\documentclass[trans,screen,nonacm]{acmart}
\AtBeginDocument{%
  \providecommand\BibTeX{{%
    \normalfont B\kern-0.5em{\scshape i\kern-0.25em b}\kern-0.8em\TeX}}}

\usepackage[utf8]{inputenc}
\usepackage[T1]{fontenc}
\usepackage{microtype}
\usepackage[ruled, linesnumbered]{algorithm2e}
\usepackage{fancyhdr,graphicx,amsmath,eqnarray}
\usepackage{caption} 
\usepackage{xcolor}
\usepackage{graphicx}
\usepackage{balance}
\usepackage{subfigure}
\usepackage{multirow}
\usepackage{tikz}
\usepackage{pgfplots}
\usepackage{filecontents}
\usepackage{threeparttable}
\usepackage{enumitem}
\pgfplotsset{compat=1.13}
\usetikzlibrary{shapes, shapes.multipart, arrows.meta, positioning, matrix, calc, fit, decorations.pathreplacing, pgfplots.dateplot}
\usepgfplotslibrary{dateplot, statistics}

\newcommand{\etal}{{\em et al.}\xspace}
\newcommand{\ie}{{\em i.e.},\xspace}
\newcommand{\eg}{{\em e.g.},\xspace}
\newcommand{\rmg}{\textsf{Muffin}\xspace}
\newcommand{\base}{\textsf{Muffin-UT}\xspace}

\begin{document}

\title[\rmg: Testing Deep Learning Libraries via Neural Architecture Fuzzing]{\rmg: Testing Deep Learning Libraries \\ via Neural Architecture Fuzzing}


\author{Jiazhen Gu}
\affiliation{%
  \department{}
  \institution{School of Computer Science, Fudan University}
  \institution{Shanghai Key Lab. of Intelligent Information Processing}
  \city{Shanghai}
  \country{China}}
  
\author{Xuchuan Luo}
\affiliation{%
  \department{School of Computer Science}
  \institution{Fudan University}
  \institution{Shanghai Key Lab. of Intelligent Information Processing}
  \city{Shanghai}
  \country{China}}

\author{Yangfan Zhou}
\authornote{Yangfan Zhou is the corresponding author.}
\affiliation{%
  \department{School of Computer Science}
  \institution{Fudan University}
  \institution{Shanghai Key Lab. of Intelligent Information Processing}
  \city{Shanghai}
  \country{China}}
  
\author{Xin Wang}
\affiliation{%
  \department{School of Computer Science}
  \institution{Fudan University}
  \institution{Shanghai Key Lab. of Intelligent Information Processing}
  \city{Shanghai}
  \country{China}}


\begin{abstract}
Deep learning (DL) techniques are proven effective in many challenging tasks, and become widely-adopted in practice. However, previous work has shown that DL libraries, the basis of building and executing DL models, contain bugs and can cause severe consequences. Unfortunately, existing testing approaches still cannot comprehensively exercise DL libraries. They utilize existing trained models and only detect bugs in model inference phase. In this work we propose \rmg to address these issues. To this end, 
\rmg applies a specifically-designed model fuzzing approach, which allows it to generate diverse DL models to explore the target library, instead of relying only on existing trained models. \rmg makes differential testing feasible in the model training phase by tailoring a set of metrics to measure the inconsistencies between different DL libraries. In this way, \rmg can best exercise the library code to detect more bugs.
To evaluate the effectiveness of \rmg, we conduct experiments on three widely-used DL libraries. The results demonstrate that \rmg can detect 39 new bugs in the latest release versions of popular DL libraries, including Tensorflow, CNTK, and Theano.

\end{abstract}


\maketitle

\section{Introduction}
\label{sec:intro}
Deep learning (DL) techniques have been proven effective in many specific tasks, such as image recognition~\cite{DBLP:conf/cvpr/HeZRS16}, video understanding~\cite{DBLP:conf/cvpr/WuF0HKG19} and machine translation~\cite{DBLP:conf/acl/WangLXZLWC19}. As a result, it becomes a trend to include DL-based functionality into traditional software design. DL systems (\ie software systems based on DL techniques) have been widely adopted in various domains in practice, \eg self-driving cars~\cite{DBLP:journals/array/GuptaAGK21}, virtual assistants~\cite{DBLP:conf/ccwc/KepuskaB18}, and software operations~\cite{DBLP:conf/kbse/0003HL0HGXDZ19,DBLP:conf/sigsoft/GuWWZLKZYSXQLLZ20, DBLP:conf/kbse/ChenZHLZHKGXDZ20}.
However, DL systems are shown to be lack of robustness, and thus cause real-world accidents. For instance, a Tesla driver was killed in self-driving mode that failed to brake the car in 2016 ~\cite{url:tesla2016}, and an Uber autonomous driving car killed a pedestrian in 2018~\cite{url:uber2018}. Errors/defects in DL systems can cause severe consequences, and even jeopardize human lives. Therefore, it is a critical task to test DL systems before deploying them in real production scenarios.

Unfortunately, how to test a DL system still remains an open challenge to the software engineering community. Many recent approaches focus on testing its core component, the DL model, \ie a deep neural network trained with a set of training data. Extensive work aims at improving the robustness of DL models via generating adequate test cases, \eg adversarial inputs or corner cases~\cite{DBLP:conf/icse/TianPJR18, DBLP:conf/kbse/ZhangZZ0K18}. Many studies also focus on designing criteria to measure the test adequacy~\cite{DBLP:conf/sosp/PeiCYJ17, DBLP:conf/icse/KimFY19}. 

However, the execution of DL models relies on their back-end libraries (\ie DL libraries). Even with a correct model design, the outputs can be wrong if the underlying library contains bugs. Specifically, DL libraries provide high-level interfaces of the underlying various computation implementations (\eg matrix transformation, gradient calculation and weight update) over hardware infrastructure (\eg CPU and GPU).
Bugs in DL libraries can inevitably cause unexpected outputs, or even fatal failure of DL systems~\cite{DBLP:conf/sigsoft/IslamNPR19}. But one may tend to blame the DL model design, instead of its underlying library,  when debugging \cite{DBLP:conf/icse/PhamLQT19}, incurring more difficulty to the process. Hence, it is critical to investigate how to test DL libraries.

Recent efforts (\ie CRADLE~\cite{DBLP:conf/icse/PhamLQT19} and LEMON~\cite{DBLP:conf/sigsoft/WangYCLZ20}) on DL library testing focus on the {\em inference phase} of DL models. 
They adopt differential testing~\cite{DBLP:conf/icse/GulzarZH19} to detect bugs, by comparing the inference results
of existing, already-trained DL models with different DL libraries.
Specifically, CRADLE directly use such models as test inputs, while LEMON further mutates such models as test inputs. 
However, even with these approaches, bugs still exist in DL libraries, as we have found in this work. 
The key reason is that they rely on the inference phase of already-trained models, which cannot exercise the library codes comprehensively. Such already-trained models typically involve only a small set of DL library functions. Moreover, DL libraries also play an important role in the model {\em training phase}, \eg the library codes for back propagation~\cite{DBLP:journals/nn/Hecht-Nielsen88a}. These library codes also cannot be exercised as well. But bugs in these codes can cause incorrect training results, \ie wrong resulting models. 

Unfortunately, solving these concerns is a challenging task. 
First, it is hard, if not infeasible, to obtain tremendous, diverse already-trained models to comprehensively exercise library codes. Mutations based on such existing models also cannot solve this problem as they inherit the model structures, limiting the exploration of library functions.
Moreover, as test oracles are not available generally, existing approaches~\cite{DBLP:conf/icse/PhamLQT19, DBLP:conf/sigsoft/WangYCLZ20} resort to differential testing, based on comparing the model outputs with different DL libraries. However, such outputs are not existing in the training phase, incurring a huge challenge to applying  differential testing.

In this work, we propose \rmg, a fuzzing-based approach to test DL libraries with high functionality coverage. Instead of relying on already-trained models, \rmg obtains diverse test inputs (\ie models) with an automatic model generation algorithm. It formulates model structure as a Directed Acyclic Graph (DAG), based on which it builds a model layer by layer with an aim to achieve high functionality coverage of DL libraries.
To perform differential testing, \rmg relies on data trace analysis in the training phase. In particular, we divide the model training phase into three different stages (\ie forward calculation, loss calculation and gradient calculation), and accordingly design a set of metrics on the data traces to measure the consistency of results by different DL libraries. Inconsistencies can thus indicate potential bugs. 

We apply \rmg to test 15 release versions of three widely-used DL libraries, \ie TensorFlow~\cite{url:TensorFlow}, CNTK~\cite{url:CNTK}, and Theano~\cite{team2016theano}. 
\rmg detects 39 new bugs (including 21 crash bugs) in the latest release versions of these libraries. Extensive experiments based on 6 popular datasets show that compared with existing approaches, \rmg is capable of detecting more inconsistencies within a comparable testing time.
Furthermore, we investigate the benefit of our model generation method through comparing \rmg with layer-by-layer testing. The results show that \rmg is capable of detecting more inconsistencies and crashes. Our experiments prove the effectiveness of \rmg.

\rmg contributes to the software testing art in the following three aspects: 
\begin{itemize}[topsep=4pt]
    \item We propose \rmg, a DL library testing approach based on a novel DL model fuzzing method, which can exercise DL library functionalities more comprehensively.
    
    \item We make differential testing feasible in testing the model training phase, by proposing a data trace analysis method to detect inconsistencies between different test targets. 
    
    \item We implement our ideas as an open-source available software tool \rmg, which can facilitate real-world DL library testing tasks, as well as further follow-up research.
    
    \item We conduct an extensive study on 15 versions of three widely-used DL libraries. The results show that \rmg can detect 39 new bugs, which cannot be detected by previous methods.
\end{itemize}

The rest of paper is organized as follows. Section~\ref{sec:background} introduces background knowledge about DL model and DL library. Section~\ref{sec:approach} elaborates the design and implementation details of \rmg. We demonstrate the experimental setup in Section~\ref{sec:evaluation}, and analyze the results in Section~\ref{sec:results}. Further discussion is provided in Section~\ref{sec:discussion}. We introduce related work in Section~\ref{sec:related} and conclude the paper in Section~\ref{sec:conclusion}.
\section{Background}
\label{sec:background}
\subsection{Deep Learning Model}
\label{sec:background:model}
DL models are designed to automatically draw statistical rules from training data~\cite{DBLP:books/daglib/0040158}. A DL model typically consists of a number of neurons with a layered, connected structure. The neurons between layers are connected with links. Different links are associated with different weights, which are obtained through training with input data. Each layer conducts a specific kind of transformation (\eg convolution and pooling) for the input data with specific weights. In particular, the same layer can be adopted multiple times in a DL model, which has different weight values on the links and thus produces diverse results. 

\begin{figure}
    \centering
    \includegraphics[width=.75\linewidth]{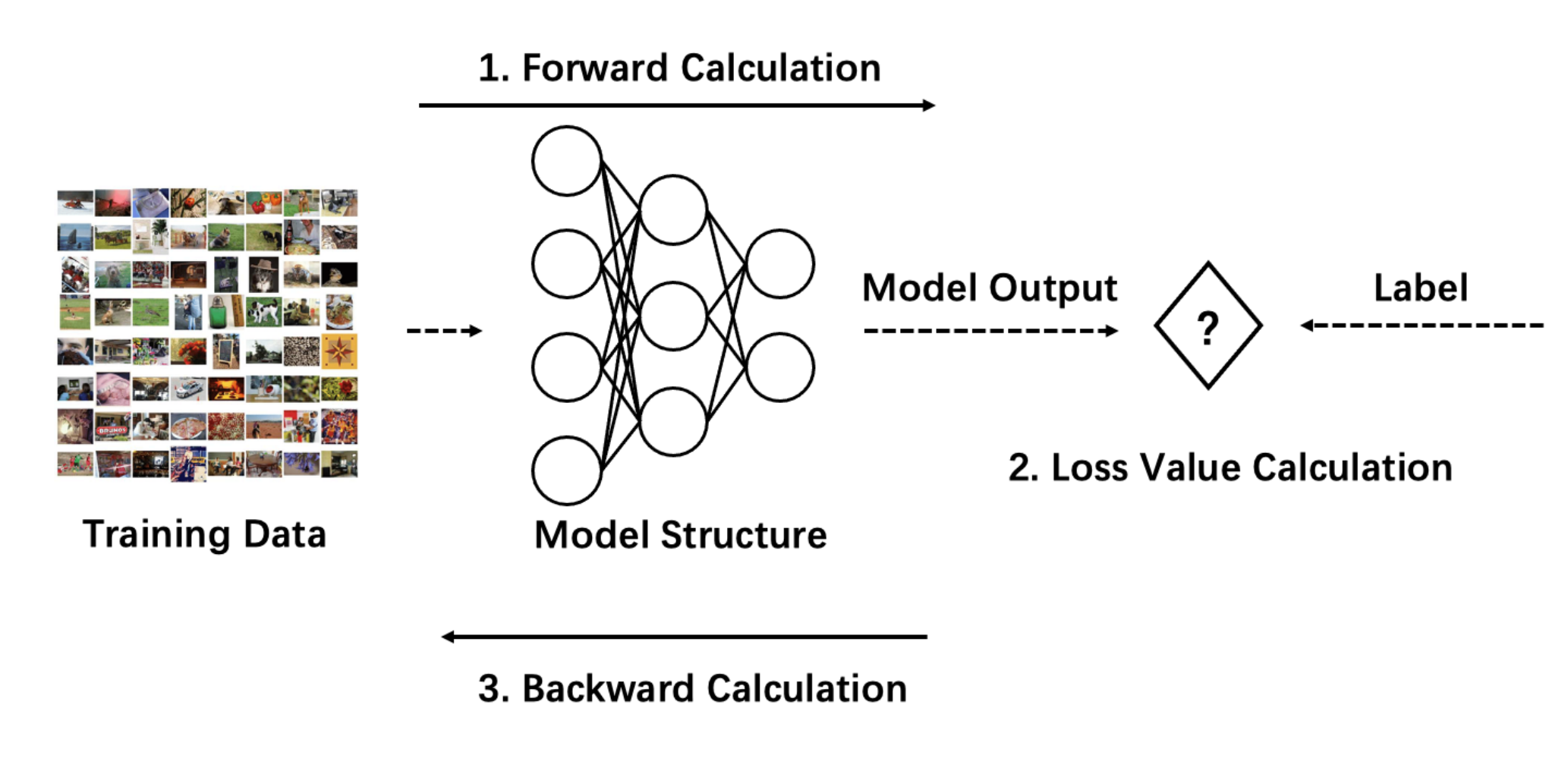}
    \vspace{-12pt}
    \caption{The training phase of a DL model}
    \label{fig:DL_training}
    \vspace{-6pt}
\end{figure}

Essentially, a developer would design the architecture of a DL model such as the types of layers, how layers are connected and the loss function. Then the training process of a DL model is to find the appropriate weight values, so that the outputs can best produce expected results. The training phase typically consists of a huge amount of repeated training steps. Figure~\ref{fig:DL_training} outlines the process of a single training step, which can be divided into three stages: 
\begin{itemize}
    \item \textit{Forward Calculation (FC)}: Given a batch of training cases, the model conducts specific calculations according to the layer types and get the corresponding outputs. 
    \item \textit{Loss Calculation (LC)}: The model calculates the value of the predefined loss function, which  measures the differences between the model outputs and the ground-truth labels.
    \item \textit{Backward Calculation (BC)}: According to the value of the loss function, the model calculates the gradients of each neuron and updates the corresponding weight values from the output layer to the input layer.
\end{itemize}
Such training steps continue until the weights converge, \ie the performance of the model cannot be further improved. 

The weight values determine how the DL model processes the input to generate output. Thus, the performance of a DL model, \ie whether it can produce correct results, is largely determined by the weights. Since the weight values are obtained through the training process, 
it is critical to detect bugs in model training phase. However, existing work only focuses on detecting bugs in the inference phase~\cite{DBLP:conf/icse/PhamLQT19, DBLP:conf/sigsoft/WangYCLZ20}. 

\subsection{Deep Learning Library}

\begin{figure}[t]
    \centering
    \includegraphics[width=.55\linewidth]{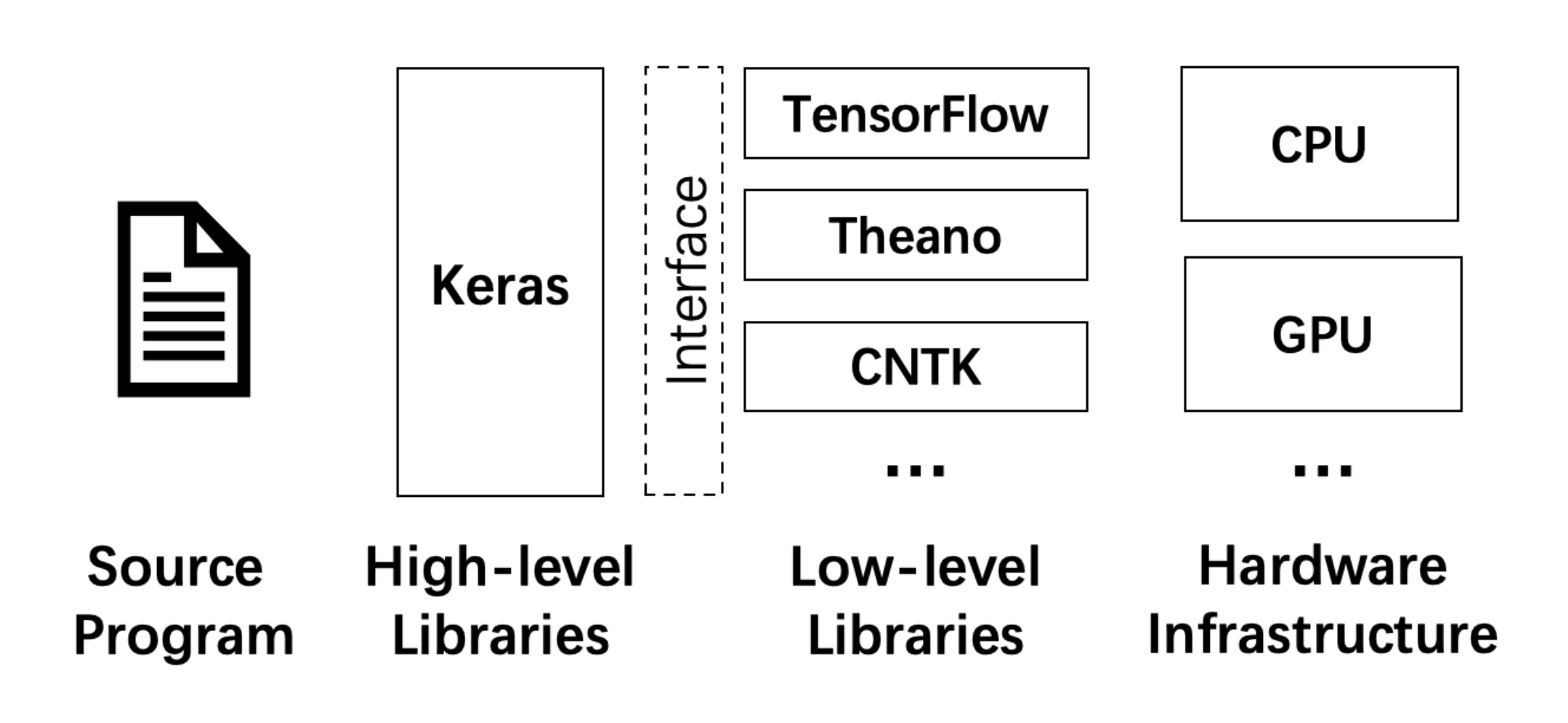}
    \caption{The structure of DL libraries}
    \label{fig:DL_lib}
    \vspace{-12pt}
\end{figure}

Figure~\ref{fig:DL_lib} shows the structure of DL libraries. There are two tiers of libraries (\ie high-level and low-level). In general, developers implement the source programs with high-level library APIs, which invoke the algorithms implemented in low-level libraries. Different low-level libraries are based on different infrastructures, \eg CPU, GPU and Tensor Processing Unit (TPU)~\cite{url:TPU}, thus may have different implementations for the same algorithm specification. On the other hand, high-level DL libraries can hide the differences between low-level libraries and provide a consistent abstraction to facilitate DL model development.

Keras~\cite{url:Keras} is one of the most popular high-level DL libraries that has been widely used in various domains~\cite{DBLP:conf/icassp/ChoiFSC17, DBLP:journals/bmcbi/HungGRLRNMNLFRD20, DBLP:conf/iccsa/MuralikrishnaVS20}. Keras generally runs on top of three low-level libraries, \ie TensorFlow, CNTK, and Theano, which cover most of the widely-used libraries. Developers implement source programs by calling APIs provided by Keras, which invoke the assigned backend low-level library to execute the computation.


Specifically, implementing a DL model using Keras mainly contains three parts: loading the data, defining the model architecture, and training the model with the data. It is worth noting that while the training process includes complicated calculations (\ie FC, LC and BC), it can be simply implemented via calling the ``model.fit()'' function provided by Keras. 

A high-level library is relatively simple, which glues the functions in a low-level library that provide concrete, complicated computation. Low-level libraries are not bug free and they are also not easy to be tested, due to their complication. Similar to existing work~\cite{DBLP:conf/icse/PhamLQT19, DBLP:conf/sigsoft/WangYCLZ20}, we focus on testing low-level libraries, \eg TensorFlow, CNTK, and Theano. We adopt Keras as the high-level library. Our target is to test the DL library codes involved in the model training phase with high functionality coverage. Specifically, DL libraries contain many auxiliary codes for various tasks such as profiling and hardware adaptation, rather than learning-related ones. Like existing tools to test DL libraries, we focus only on learning-related APIs. In this paper, we use functionality coverage as the coverage metric, which refers to the percentage of the invoked APIs in all the pre-defined, learning-related APIs we considered.

\subsection{Challenges}
In order to perform comprehensive DL library testing (\eg test the library codes involved in model training), there are two main challenges. First, it is difficult to obtain a set of DL models as testing inputs that cover most library APIs. A DL model has a layered, connected structure, which hinders the adoption of traditional test input generation approaches. Furthermore, many APIs in DL libraries have specific usage scenarios, \eg {\em Convolution Layer} for image processing tasks, {\em Recurrent Layers} for text processing and various activation functions (\eg ReLU~\cite{DBLP:conf/icml/NairH10} and leaky-ReLU~\cite{DBLP:journals/corr/XuWCL15}). Due to such complication, it is non-trivial to obtain a set of well-trained models to achieve high functionality coverage.

The second challenge is the test oracle in model training phase. Existing approaches~\cite{DBLP:conf/icse/PhamLQT19, DBLP:conf/sigsoft/WangYCLZ20} utilize differential testing based on the model outputs with different DL libraries. Unfortunately, as we have discussed, DL models learn the weight values through training. Therefore, the model outputs not exist in the training phase, causing existing differential testing methods infeasible. 

Next, we introduce our approach, \rmg, which is designed to address the above two challenges.
\begin{figure*}[ht]
    \centering
    \includegraphics[width=.99\textwidth]{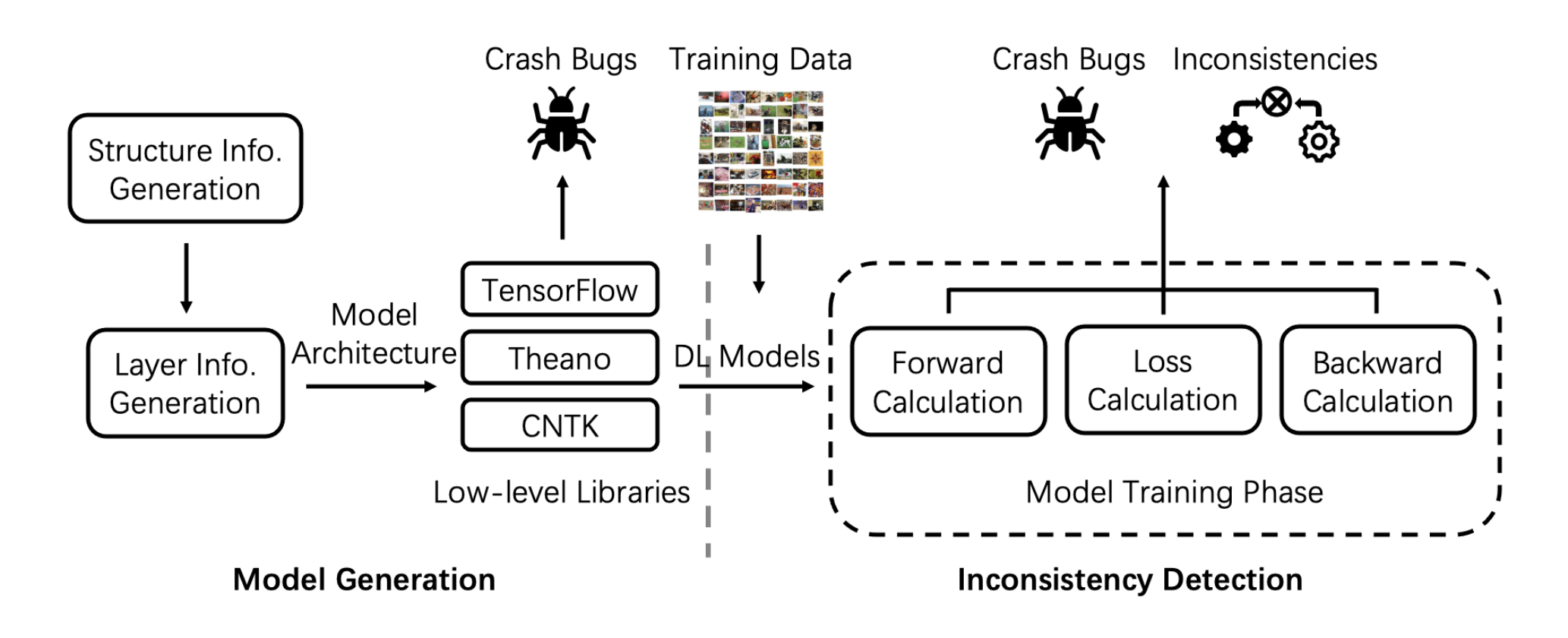}
    \vspace{-12pt}
    \caption{Overview of \rmg}
    \label{fig:overview}
\end{figure*}

\section{Approach}
\label{sec:approach}

\subsection{Overview}
In this work, we propose \rmg, a novel approach to perform comprehensive DL library testing, \ie test the library codes related to model training with high functionality coverage. Figure~\ref{fig:overview} presents the overview of \rmg, which is specifically tailored to solve the two design challenges. 

To obtain diverse DL models, we propose a fuzzing-based model generation method. In contrast to existing methods that adopt manually-designed models, the proposed model generation approach allows \rmg to exercise the target library with tremendous, diverse models. 
Specifically, we divide the model architecture into two parts: structure information (\ie how layers are connected) and layer information (\ie what layer types are used). Through formulating the structure information of a DL model as a DAG, \rmg first generates DAGs as the structure information, and then utilizes a greedy layer selection algorithm to generate the layer information. In this way, \rmg can generate diverse DL models (Section~\ref{sec:approach:generation}). 

To conduct differential testing, \rmg performs data trace analysis in the model training phase. In particular, \rmg profiles the data traces from different training stages (\ie FC, LC and BC). It then detects the inconsistencies of different libraries based on a set of proposed metrics, which measures the output variance of consecutive layers. (Section~\ref{sec:approach:detection}).

\subsection{Model Generation}
\label{sec:approach:generation}
As discussed in Section~\ref{sec:background:model}, a DL model has a layered structure with connections between layers. In order to generate a set of diverse DL models to explore library codes, we need to decide what types of layers are used in a model, as well as how these layers are connected. Unfortunately, simply selecting a series of layers and stacking them together can easily cause model failure. For example, the ``Add'' layer is used to add a list of inputs. If only one input is fed to this layer, the model generation would fail. Besides, the inputs of ``Add'' layer should also have the same shape to avoid failure generation. Therefore, we design a top-down generation algorithm, which first generates the \textit{structure information} (\ie the topology of how layers are connected in the model), followed by the generating of the \textit{layer information} (\ie specific layer types adopted in the model).

\subsubsection{Structure Information Generation}
Given a set of inputs, a DL model performs specific computation layer by layer, so as to yield the outputs. Therefore, the computation flow of a DL model can be abstracted as a DAG. Specifically, every vertex in the DAG represents a layer, and every edge between two vertices represents a link between the corresponding layers in the original model. Such an abstraction method is also applied in current model representation. For example, TensorFlow uses a DAG to represent the computational graph of a DL model~\cite{DBLP:conf/osdi/AbadiBCCDDDGIIK16}. Therefore, we utilize a DAG to represent the \textit{structure information} of a DL model.

\begin{figure}
    \subfigure[Chain structure with skips]{
        \begin{minipage}[t]{0.30\linewidth}
        \centering
        \includegraphics[width=\linewidth]{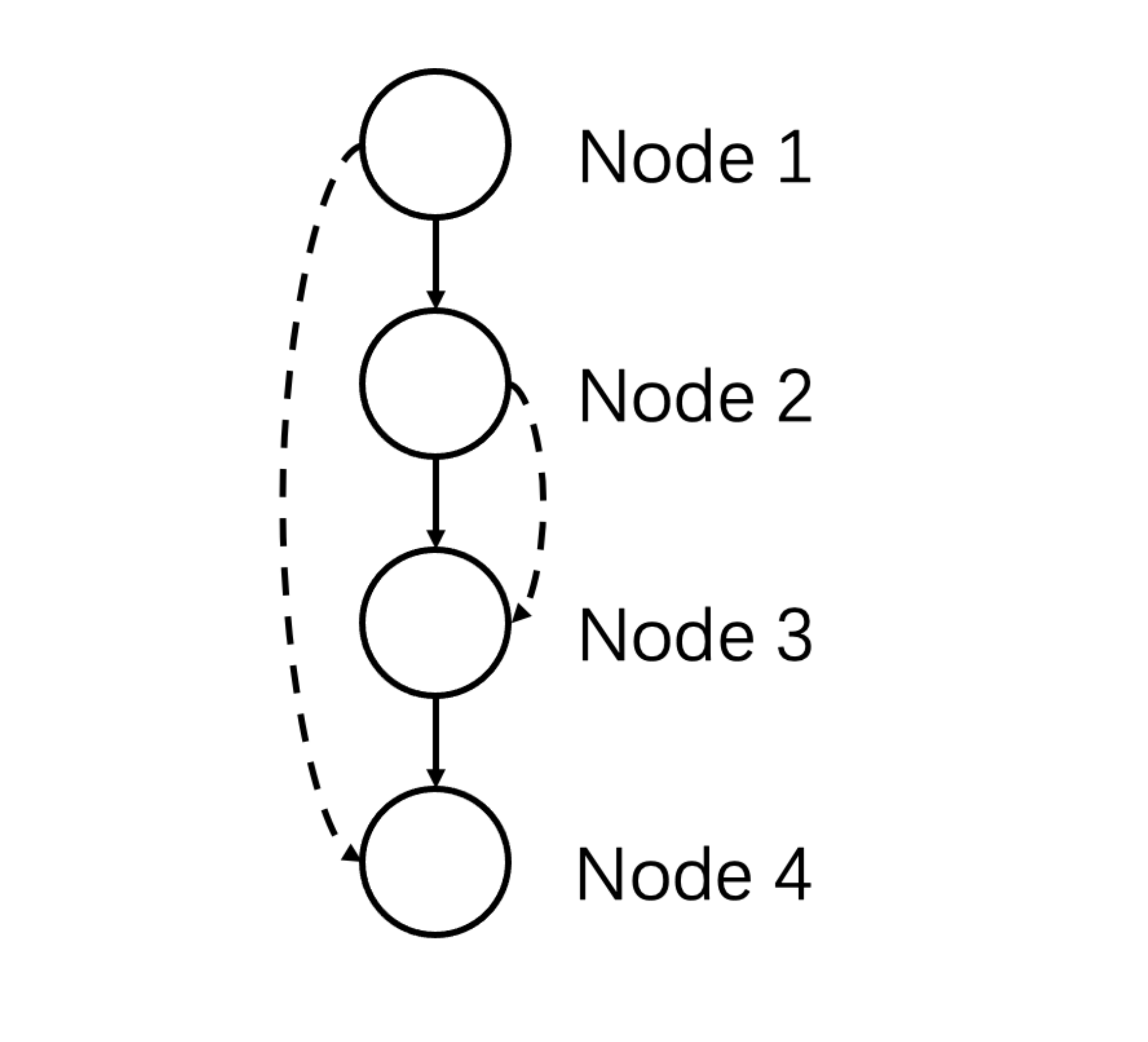}
        \label{fig:template:chain}
        \end{minipage}
    }
    \subfigure[Cell-based structure]{
        \begin{minipage}[t]{0.30\linewidth}
        \centering
        \includegraphics[width=\linewidth]{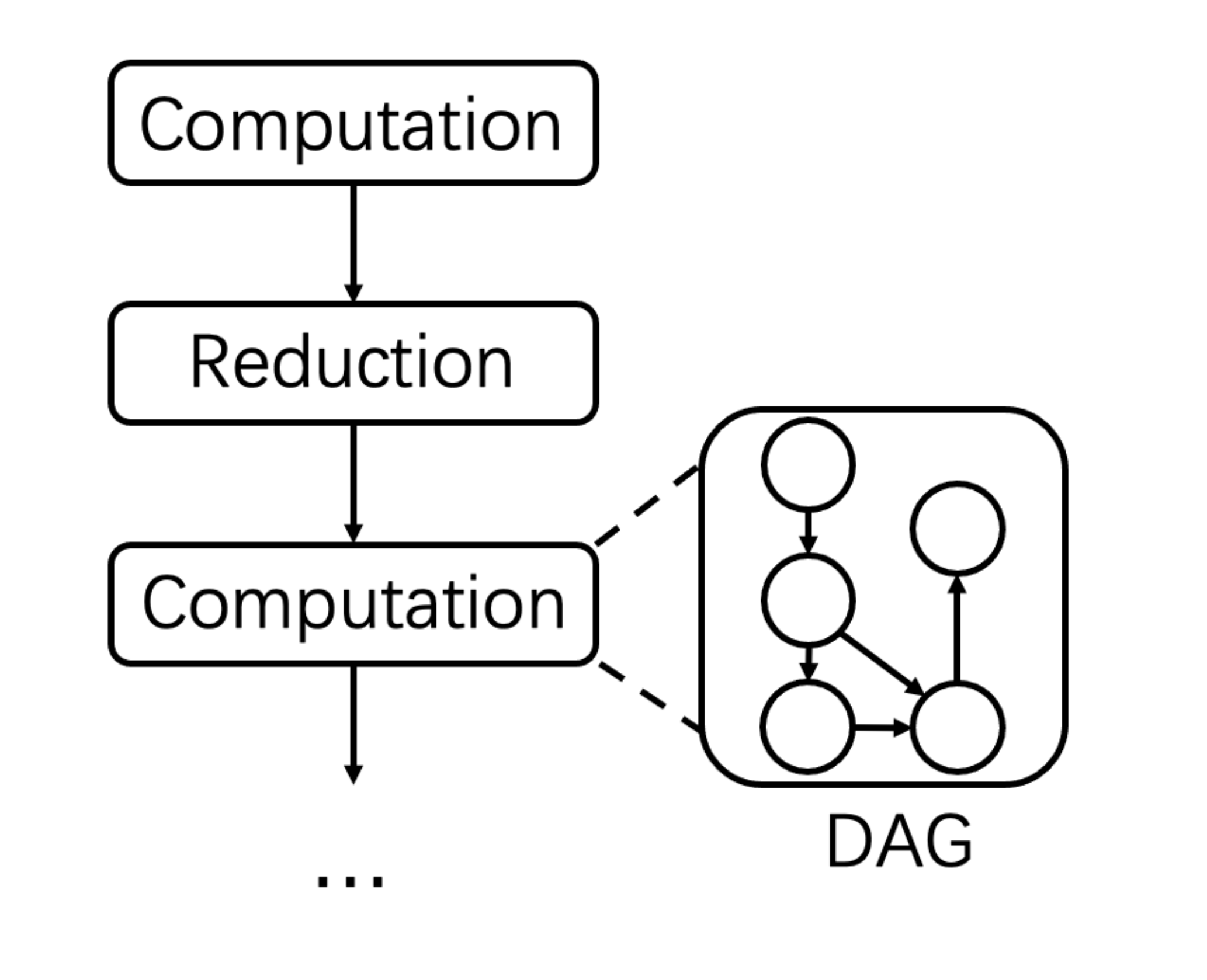}
        \label{fig:template:cell}
        \end{minipage}
    }
    \caption{Examples of DL model structure templates}
    \label{fig:template}
    \vspace{-12pt}
\end{figure}

Although it is not difficult to generate a DAG, the corresponding model structure may be too simple or too complicated, which is rarely used in practice. Inspired by recent studies in Neural Architecture Search (NAS) ~\cite{DBLP:journals/jmlr/ElskenMH19}, that targets on automating the design of model architectures, we summarize two model structure templates, as shown in Figure~\ref{fig:template}. Specifically, Figure~\ref{fig:template:chain} shows the chain structure with skips. Chain-structured architecture is the simplest example of the model structure topology. Through permitting arbitrary skip connections between nodes, this template can cover many commonly-used DL models (\eg fully-connected networks, VGG~\cite{DBLP:journals/corr/SimonyanZ14a} and DenseNet~\cite{DBLP:conf/cvpr/HuangLMW17}). On the other hand, the cell-based structure in Figure~\ref{fig:template:cell} builds upon the observation that many specifically-designed model architectures consist of repetitions of fixed structures~\cite{DBLP:journals/corr/abs-1905-01392}, \eg ResNet~\cite{DBLP:conf/cvpr/HeZRS16}. Each cell in the structure is a small DAG that conducts a specific transformation, \eg the computation cell in Figure~\ref{fig:template:cell} contains computation layers, while the reduction cell is used for downsampling. It is worth noting that originally the same cells (\eg computation cells) should have the same DAG. Since our target is generating diverse structures instead of finding the architecture with the best performance, we remove this restriction in the proposed template (\ie same cells may have different DAGs). In addition, we also guarantee that the generated DAG has only one vertex whose in-degree is 0 as the input layer, one vertex whose out-degree is 0 as the output layer. There is also no isolated vertex in the generated DAG. In this way, \rmg generates a DAG as the model structure information.

\subsubsection{Layer Information Generation}
Given the generated structure information, we need to refine the layer information, \ie determine the specific layer type for each vertex in the DAG. As discussed before, stacking layers without guidance can easily cause model failure. Specifically, there are two types of restrictions when selecting layers. The first restriction is the \textit{input number} restriction. In particular, most layers (\eg 
``Convolution'') are SI (Single-Input) layers, while some layers (\eg ``Concatenation'') are MI (Multiple-Input) ones. If more than one input is fed to an SI layer, this layer can only process one of the inputs, leading to the existence of invalid connections between layers. The corresponding DAG thus is equivalent to the DAG without the invalid connections, which lowers the DAG diversity. Therefore, we have to choose the proper layer according to the number of inputs. Considering that an edge in a DAG represents the data flow direction, we can determine the number of inputs that the corresponding layer takes based on the {\em in-degree} of the vertex.

The second restriction is the {\em input/output shape} restriction. Specifically, MI layers require the inputs to have the same shape in specific axis(es) so as to conduct the transformation properly, \eg the inputs of ``Concatenation'' layer should have the same shape except for the concatenation axis. Therefore, before feeding the inputs to an MI layer, we adopt additional ``Reshaping'' layers to reshape the inputs into the same shape. In addition, the input shape of the input layer (\ie the vertex with 0 in-degree), and the output shape of the output layer (\ie the vertex with 0 out-degree) need to be properly set according to the training data and the task type (\eg classification, regression). We still resort to the ``Reshaping'' layer to reshape the output size, while the input shape is directly set based on the shape of input data.

Furthermore, in order to increase the diversity of the generated models, intuitively, we should give a larger chance to the layer that is rarely used before. Based on this intuition, we design a layer selection procedure based on Fitness Proportionate Selection~\cite{DBLP:journals/complexity/Fogel97}. Specifically, for a specific layer $l$, \rmg records the number of times that $l$ has been selected to construct a model, denoted as $c$. Then \rmg calculates $s = \frac{1}{c+1}$ as the score for $l$. Based on $s$, the probability that $l$ is selected among all layer types can be calculated as follows:

\begin{equation}
    p = \frac{s}{\sum^{r}_{k=1}s_k}
\end{equation}

where $r$ is the total number of possible layers. Since we divide layer types into two categories (\ie SI and MI), score $s$ and probability $p$ are calculated based on the layers belonging to the same category. In this way, \rmg generates the layer information via selecting a specific layer for each vertex in the DAG. 
A DL model can thus be constructed according to the generated structure information and layer information.


\subsubsection{Entire Algorithm}
\SetAlgoVlined
\SetKw{Continue}{continue}
\begin{algorithm}[t]
    \KwIn{
        $N_m$: Number of generated models \\
        \quad\quad\quad $MAX_c$: Maximum number of cells in a model \\
        \quad\quad\quad $MAX_v$: Maximum number of vertices in a DAG \\
        \quad\quad\quad $L_i$: Input shape \\
        \quad\quad\quad $L_o$: Output shape
    }
    \KwOut{$M$: A set of generated models }

    $M \leftarrow \emptyset$\;
    \While{$Size(M) < N_m$}{
        \tcc{select a template and generate Structure Information $SI$}
        $p \leftarrow Random(0,1)$\;
        \eIf{$p < 0.5$}{
        $N_v \leftarrow RandomInt(1, MAX_v)$\;
        $SI = CreateChainDAG(N_v)$\;
        }{
        $N_c \leftarrow RandomInt(1, MAX_c)$\;
        $G \leftarrow \emptyset$\;
        \For{$i$ from 1 to $N_c$}{
            $G_i = RandomDAG()$\;
            $G \leftarrow G \cup \{G_i\}$\;
        }
        $SI = CreateCellDAG(N_c, G)$;
        }
        $LI \leftarrow \emptyset $\tcc*{Layer Information}\
        \tcc{generate model according to DAG}
        \ForEach{node $j$ in $TopologicalSequence(SI)$}{
            \uIf{the in-degree of node $j$ is 0}{
                $LI \leftarrow LI \cup \{SetInputLayer(L_i, j)\}$\;
            }
            \uElse{
                $P_j \leftarrow GetAllDirect Predecessors(j)$\;
                \uIf(\tcc*[f]{SI layer}){$Size(P_j) == 1$}{
                    $shape \leftarrow GetShape(P_j[0])$\;
                    $LI \leftarrow LI \cup \{SetSILayer(shape, j)\}$\;
                    $UpdateSIScore()$
                }
                \uElse(\tcc*[f]{MI layer}){
                    $shape \leftarrow RandomShape()$\;
                    $LI \leftarrow LI \cup ReshapingLayers(P_j, shape)$\;
                    $LI \leftarrow LI \cup \{SetMILayer(shape, j)\}$\;
                    $UpdateMIScore()$
                }
            }
            \uIf{the out-degree of node $j$ is 0}{
                $LI \leftarrow LI \cup SetOutputLayer(j, L_o)$\;
            }
        }
        $m \leftarrow ConstructModel(SI, LI)$\;
        $M \leftarrow M \cup \{m\}$\;
    }
    \Return $M$\;
    \caption{Model Architecture Generation}
    \label{alg:model_generation}
\end{algorithm}
We formally describe our fuzzing-based model generation method in Algorithm~\ref{alg:model_generation}. This algorithm takes 5 parameters, where $N_m$ is the total number of models to generate, serving as the terminating condition; $MAX_c$ and $MAX_v$ are parameters to control the size of DAG; $L_i$ and $L_o$ should be manually set according to the input data and target task. Lines 2-33 iteratively generate a set of DL models. Specifically, lines 3-13 randomly choose a template and generate a DAG as the structure information. Lines 17-18 set the input layer. Lines 22-24 select SI layers for the 1 in-degree vertices, and Lines 26-29 select MI layers for the vertices with more than 1 in-degree. Lines 30-31 set the output layer. Finally, lines 32-33 construct a DL model $m$ based on the generated structure information and layer information, then adding $m$ to the result set $M$.

\subsection{Inconsistency Detection}
\label{sec:approach:detection}
In order to detect inconsistencies and perform differential testing accordingly, \rmg requires proper metrics to measure the differences between the execution results of different libraries. However, the metrics proposed by the existing work are designed only for already-trained models, which calculate the inconsistency between the ground-truth label and model outputs. Since our target is to test DL library in the training phase (\ie without a trained model), such metrics cannot be directly applied. Instead, we propose a new metric based on the variance of outputs in consecutive layers.

As demonstrated in Section~\ref{sec:background:model}, the model training phase includes repeated training steps, and each training step can be divided into three stages: FC, LC and BC. Specifically, in FC stage, the model performs calculation from input layer to output layer. In LC stage, the model calculates the value of loss function. In BC stage, the model calculates gradients from the output layer to the input layer. Resorting to dynamic analysis, we can collect the data traces of different libraries and compare the differences.

In particular, we utilize the {\em Functional API} mechanism provided by Keras to collect dynamic traces. More specifically, we profile the results produced by every layer in FC stage, the loss value in LC stage, and the gradient value of each layer in BC stage. Based on the dynamic trace, \rmg gradually compares the values of layer outputs, the loss, and the gradients, so as to detect the suspect behavior of specific layers.

However, due to normal uncertain factors such as floating-point deviation~\cite{DBLP:journals/csur/Goldberg91}, we cannot determine whether a value difference is caused by potential bugs or normal factors. Specifically, there are many small deviations (less than $10^{-6}$) in layer outputs, which may be gradually amplified or reduced, \eg by pooling or activation functions. We consider that normal factors would only lead to slight layer output difference, \ie if the differences of layer outputs change dramatically, it indicates a suspicious behavior. Therefore, instead of comparing the outputs from one single layer, we consider the difference-changes between two consecutive layers. Only when the deviation is amplified would \rmg consider inconsistency.
Since outputs from different layers may have different shapes, we first use the following Chebyshev distance (\ie $L_{\infty}$ distance)~\cite{cantrell2000modern} to measure the difference of the outputs from the same layer.

\begin{equation}
    D(X,Y) = \max_m(|x_m-y_m|)
\end{equation}

In the above equation, $X$ and $Y$ are two tensors (\ie output of a layer is typically a high-dimensional tensor), while $x_p$ and $y_p$ are elements in $X$ and $Y$, respectively. Chebyshev distance defines that the distance between two tensors is the greatest of their differences along any coordinate dimension. In this way, we can avoid the influence of different tensor shapes from different layers when measuring the differences.

We now describe the inconsistency detection procedure of \rmg. For brevity, we denote $n$ as the total number of layers, $l_i, i\in[1,n]$ as the $i^{th}$ layer, $O^i_j$ and $O^i_k$ as the outputs of $l_i$ using library $j$ and $k$, respectively. $P(i)$ denotes the set of layers that are direct predecessors of $l_i$ in the DAG, \ie each layer in $P(i)$ is $l_i$'s previous layer.

In FC stage, \rmg compares the differences of the output tensors from $l_i$ and its predecessors $l_p$. If the difference of $l_p$ is smaller than $\epsilon$, while the difference of $l_i$ is larger than a user-defined threshold $t$, then \rmg determine that an inconsistency is detected in $l_i$. The inconsistency layers detected in FC stage can be formally defined as follows:

\begin{align*}
    Inc\_FC = \{l_i, i \in [1, n] \mid (D(O^i_j, O^i_k) > t) \land \\ (D(O^p_j, O^p_k) < \epsilon, p \in P(i)) \} 
\end{align*}

In LC stage, the model calculates loss value based on the results from the output layer. To avoid the transmission of errors, \rmg only performs inconsistency detection in LC stage when the difference of model outputs is smaller than $\epsilon$. It is worth noting that a small difference in model outputs does not mean that there is no inconsistency in middle layers. Large difference could be masked due to the existence of downsampling layers such as pooling.
Since the result of a loss function is a number, we directly compare the absolute difference as follows:


\begin{align*}
    Inc\_LC = \{L \mid ((|LO_j - LO_k| > t) \lor (|LG_j - LG_k| > t)) \land \\ (D(O^n_j, O^n_k) < \epsilon) \}
\end{align*}

In the above equation, $L$ denotes the loss function, $LO_j$ and $LO_k$ are the output results of $L$, $LG_j$ and $LG_k$ are the gradient results of $L$, $O^n_j$ and $O^n_k$ are the model outputs, using library $j$ and $k$.

In BC stage, the model calculates gradients to update weights from the output layer to the input layer. Similarly, \rmg only conducts inconsistency detection if the difference in loss function is smaller than $\epsilon$. We formulate the inconsistency detection in BC stage as follows:

\begin{align*}
    Inc\_BC = \{l_i, i \in [1, n] \mid (D(G^i_j, G^i_k) > t) \land \\ (D(G^s_j, G^s_k) < \epsilon, s \in S(i)) \} 
\end{align*}

where $S(i)$ denotes the set of layers that are direct successors of $l_i$; $G^i_j$ and $G^i_k$ denote the gradient result of $l_i$ using different libraries. Especially, the successor of the output layer is the loss function.

\section{Evaluation Setup}
\label{sec:evaluation}
In the evaluation, we evaluate the performance of \rmg through answering the following research questions.
\begin{itemize}[leftmargin=*]
    \item \textbf{RQ1:} How does \rmg perform in detecting bugs in DL libraries?
    \item \textbf{RQ2:} Can \rmg achieve better performance compared to other methods?
    \item \textbf{RQ3:} How do the different parameter settings affect the performance of \rmg?
\end{itemize}

\subsection{Libraries and Datasets}
\subsubsection{Libraries}
We use three widely-used DL libraries (\ie TensorFlow, Theano, and CNTK) as the back-end low-level libraries as our testing targets, and Keras as the front-end high-level library. To sufficiently illustrate the effectiveness of \rmg, we utilize a total of 15 release versions of the three back-end libraries, and construct five experimental environments for differential testing, \ie E1-E5 in Table~\ref{table:library_versions}. In particular, in E1, Keras 2.3.1 is the latest version that supports multiple back-ends; Theano 1.0.4 and CNTK 2.7.0 are the latest versions, while TensorFlow 2.0.0 is the latest version that supported by Keras. For the sake of brevity, we use TF, TH, and CK to represent TensorFlow, Theano and CNTK in the following figures and tables.

\begin{table}[t!]
    \centering
    \small
    \caption{Versions of libraries under test}
    \renewcommand{\arraystretch}{0.7}
    \resizebox{0.55\linewidth}{!}{

        \begin{tabular}{c|c|ccc}
            \toprule
            \textbf{ID} & \textbf{Keras} & \textbf{TensorFlow} & \textbf{Theano} & \textbf{CNTK} \\ 
            \midrule
            \textbf{E1} & 2.3.1          & 2.0.0               & 1.0.4           & 2.7.0         \\
            \textbf{E2} & 2.3.1          & 1.15.0              & 1.0.3           & 2.6.0         \\
            \textbf{E3} & 2.2.4          & 1.12.0              & 1.0.2           & 2.5.0         \\
            \textbf{E4} & 2.2.4          & 1.11.0              & 1.0.1           & 2.4.0         \\
            \textbf{E5} & 2.2.4          & 1.10.0              & 1.0.0           & 2.3.0         \\ \bottomrule
        \end{tabular}


    }
    \label{table:library_versions}
    \vspace{-10pt}
\end{table}

\subsubsection{Datasets}
Our approach is not sensitive to datasets, \ie theoretically any data type can be used for testing. In order to facilitate subsequent comparative experiments with existing approaches, we selected 6 widely-used datasets in existing studies~\cite{DBLP:conf/sigsoft/WangYCLZ20}, \ie MNIST, F-MNIST, CIFAR-10, ImageNet, Sine-Wave and Stock-Price. Specifically, the first four are popular image classification datasets, while the last two are sequence datasets. In particular, Sine-Wave is the sine function value sequence, and Stock-Price is the Disneyland stock price sequence from 1997 to 2016.

\subsection{Competitors}
In order to demonstrate the effectiveness and efficiency of \rmg, we compare \rmg with the state-of-the-art approach, LEMON~\cite{DBLP:conf/sigsoft/WangYCLZ20}. LEMON performs DL library testing though mutating existing models to generate a huge amount of new test inputs. Following the evaluation setup in~\cite{DBLP:conf/sigsoft/WangYCLZ20}, we use 11 existing models (\ie AlexNet, LeNet5, ResNet50, MobileNetV1, InceptionV3, DenseNet121, VGG16, VGG19, Xception, LSTM-1, LSTM-2) as the seed models for mutation. By comparing with LEMON, we evaluate whether \rmg, based on directed test cases (\ie model) generation, can outperform LEMON in exposing bugs. Since LEMON cannot perform testing in the LC and BC stages, we only compare \rmg with LEMON through analyzing the number of inconsistencies and bugs detected in the FC stage. In addition, \rmg is designed to perform comprehensive library testing, so we also compare the functionality coverage achieved by \rmg and LEMON.

Besides, our DAG-based model generation is the core component of \rmg. Thus, it is also interesting to investigate the effectiveness of this component. To this end, we implement \base, a simplified \rmg-version method based on unit testing. \base differs from \rmg only in the model generation part. Specifically, in \base, a functional layer (\eg Conv2D) is a to-be-tested unit. \base creates models with only one functional layer, and simple reshaping layers to cope with input/output, \ie dimension transformation to match the input/output requirements of the to-be-tested layer. By comparing with \base, we show that unit testing is still inadequate to test DL libraries. \rmg, by generating diverse models, can expose bugs which are difficult to be detected by traditional approaches.

\subsection{Measurements}
\subsubsection{Number of inconsistencies}
Since the proposed approach conducts inconsistency detection in layer level during model training, an inconsistency between two low-level DL libraries means that they produce different calculation results given the same input under a specific layer. In order to eliminate duplicated inconsistencies caused by the same function, we only count the inconsistencies produced by the same layer once. In particular, for \rmg and \base, we compare the number of inconsistencies detected in different training stages, \ie FC, LC and BC, respectively. For LEMON, we only count the inconsistencies detected in FC stage. Although different inconsistencies may be the manifests of the same potential bug, more failure-triggering tests (\ie the model and input data that trigger the inconsistency) reflecting a fault in different ways provide more information for fault localization. Therefore, the number of detected inconsistencies can reflect the effects of these methods to some extent.

\subsubsection{Number of detected bugs}
Although we count the number of detected inconsistencies, it is more important to measure the number of unique bugs revealed by \rmg. Based on the voting mechanism of differential testing, we can localize the buggy layer in the library. To avoid false positives, we further check the buggy layer manually. 
Specifically, we save all intermediate layer outputs during the testing. When an inconsistency is reported, two authors check the corresponding source codes in different libraries and compare the results. If their identical layer produces different results and their implementation ideas are different, the third author will join manual inspection so as to conclude whether the report is true or false positive.

\subsubsection{Number of NaN/Crash bugs}
Besides inconsistent calculation results, bugs in DL libraries may lead to NaN (Not a Number) and crashes as well~\cite{DBLP:conf/icse/PhamLQT19}. Generating DL models that trigger NaN/crashes can also provide valuable information for identifying potential bugs. Therefore, we count the number of models with NaN or crashes generated by three methods. In particular, we only count the NaN/crash when at least one of the DL library can execute properly, \eg TensorFlow produces normal results while Theano and CNTK produce NaN. In addition, in order to avoid duplication, the NaN caused by the same layer, and crashes with the same error message are only counted once.

\subsection{Implementations}
\label{sec:evaluation:imp}
In the experiments, we let each method generate a total of 300 models, 50 for each dataset. For LEMON, we use its default parameters. For \rmg, we set the maximum number of cells (\ie $MAX_c$) to 5, and the maximum number of nodes (\ie $MAX_v$) to 30. In terms of inconsistency detection, we set the threshold $t$ to be 0.15, and $\epsilon$ to be $1e^{-5}$. This $t$ value is relatively large so as to avoid many false positives, as shown in Section~\ref{sec:results:rq3}. In addition, \rmg do not consider some layers such as ``Dropout'' and ``GaussianNoise'', so as to avoid introducing randomness and affecting the execution results.

All the experiments are conducted on the Intel(R) Core(TM) i7-6700K CPU @ 4.00GHz machine with 32GB of RAM, Ubuntu 20.04.2 LTS, and one Nvidia GTX 1080 Ti GPU. 

The implementation of \rmg is publicly available on GitHub~\footnote{https://github.com/library-testing/Muffin}. 

\section{Results and Analysis}
\label{sec:results}
\subsection{Effectiveness of Bug Detection}
\label{sec:results:rq1}
\begin{table}[t!]
    \centering
    \caption{New bugs and crashes detected by \rmg}
    \renewcommand{\arraystretch}{0.7}
    \resizebox{0.65\linewidth}{!}{
     \begin{threeparttable}

        \begin{tabular}{@{}l|cccc|c@{}}
            \toprule
            \textbf{Library}    & \textbf{FC Bug} & \textbf{LC Bug} & \textbf{BC Bug} & \textbf{NaN} & \textbf{Crash} \\ \midrule
            \textbf{TensorFlow} & 0               & 2               & 1               & 0(1)            & 1              \\
            \textbf{Theano}     & 8               & 0               & 2               & 0            & 10             \\
            \textbf{CNTK}       & 4               & 0               & 0               & 1            & 10             \\ \midrule
            \textbf{Total}      & \multicolumn{4}{c|}{18(1)}                                            & 21             \\ \bottomrule
        \end{tabular}

        \begin{tablenotes}
            \footnotesize
            \item[1] "FC Bug""LC Bug""BC Bug" respectively refer to new bugs found in Forward Calculation, Loss Calculation and Backward Calculation stages.
            \item[2] "NaN" refers to bugs related to NaN calculation.
            \item[3] The number in parentheses means the bug exists in TensorFlow2.0.0 but has been fixed in the latest version. Other bugs are all exists in the latest version.
        \end{tablenotes}

    \end{threeparttable}
    }
    \label{table:new_bugs}
    \vspace{-15pt}
\end{table}
We first investigate the effectiveness of \rmg in terms of new bugs detected in the latest versions of different libraries, \ie E1 in Table \ref{table:library_versions}. After manual analysis, \rmg detects 18 bugs in the latest version of these libraries, including 12 bugs in FC stage, 2 bugs in LC stage, 3 bugs in BC stage and 1 NaN bugs, as shown in Table \ref{table:new_bugs}. In addition, \rmg also detects 21 crash bugs, mainly from Theano and CNTK. 

In particular, for the 4 bugs detected in TensorFlow 2.0.0, we manually check whether these bugs can be reproduced in the latest version (\ie TensorFlow 2.6.0). The results show that among these bugs, 1 bug has been fixed while the other 3 bugs still exist. After reporting these bugs to the issue repository, 1 bug has been confirmed by developers.
Among the 4 bugs detected in CNTK, 1 will be fixed in the future version~\cite{url:CNTK_ops}.
We also provide bug case analysis according to different bug types.
    
\textbf{FC Bugs.} The 12 bugs detected in FC stage involve different layer types, including ``AveragePooling2D'', ``Conv1D'' in Theano, and ``LSTM'', ``DepthwiseConv2D'', ``BatchNormalization'' in CNTK. By taking the ``AveragePooling2D'' bug in Theano as an example. This bug occurs when setting the layer parameter $padding$ to ``same`` and $pool\_size$ to the same as the shape of the input tensor. By analyzing the results, we find that in this case, Theano would choose a wrong pooling location, resulting in large difference (\ie more than 13 while $t=0.15$) between the results from other libraries.

\textbf{LC Bugs.} By taking the ``BinaryCrossentropy'' bug in TensorFlow as an example. When passing parameter values $output$=[0., 1., 0.] and $target$=[0.9999999, 0.9999999, 0.0000001] to the ``BinaryCrossentropy'' loss function, theano and CNTK return a value [15.942385, 1.1920930e-07, 1.1920930e-07] while TensorFlow returns [15.333239, -0., -0.], among which the difference of the first element is not negligible. By reviewing the source code, we find that TensorFlow redundantly uses an $epsilon$ parameter to clip input values, resulting in errors. This bug has been confirmed by the developers of TensorFlow.

\textbf{BC Bugs.} By taking the ``ReLU'' bug in Theano as an example. When $0$ exists in the input tensor of $ReLU$, theano back-propagates a different gradients value, compared with TensorFlow and CNTK. This bug is caused by the wrong equal sign position of Theano, \ie $ReLU(z)=z \mid z \geq 0$ in Theano, while $ReLU(z)=z \mid z > 0$ in other libraries. Although such implementation does not affect the results in forward calculation, the implementation in Theano would let the gradient propagate to previous layers in backward calculation (which should not happen). This bug can only be detected in BC stage, proving the effectiveness of \rmg.

\textbf{NaN Bugs.} By taking a TensorFlow bug as an example. Given two $NaN$ value, the ``GlobalMaxPooling'' layer returns $-INF$, leading to the inconsistency. This bug has been fixed in the latest 2.6.0 version.

Regarding false positives, \rmg reports 19 unique inconsistencies totally, where one false positive is found. The false positive occurs in the "mean\_absolute\_percentage\_error" loss function. This function returns $100 \times mean$, which amplifies the deviation and cause the false alarm.
In addition, \rmg detects 25 crash bugs totally. Among them four are due to unsupported models \rmg generates, which can be treated as false positives. But, such false positives have clear error messages, thus can be automatically detected so as to avoid false alarms.

\begin{table}[t!]
    \centering
    \caption{Comparison of distinct voted layers}
    \renewcommand{\arraystretch}{0.7}
     \begin{threeparttable}
        \begin{tabular}{@{}l|p{0.6cm}<{\centering}|p{1.2cm}<{\centering}p{1.2cm}<{\centering}p{1.2cm}<{\centering}@{}}
            \toprule
            \textbf{Method}                    & \textbf{Lib} & \textbf{FC} & \textbf{LC} & \textbf{BC} \\ \midrule
            \multirow{3}{*}{\textbf{\rmg}}      & \textbf{TF}  & 3 (2)        & 1 (1)        & 1 (1)        \\
                                               & \textbf{TH}  & 15 (5)       & 1 (0)        & 1 (1)        \\
                                               & \textbf{CK}  & 6 (2)        & 1 (0)        & 1 (1)        \\ \midrule
            \multirow{3}{*}{\textbf{LEMON}}    & \textbf{TF}  & 2 (0)        & -            & -           \\
                                               & \textbf{TH}  & 1 (0)        & -            & -           \\
                                               & \textbf{CK}  & 1 (0)        & -            & -           \\ \midrule
            \multirow{3}{*}{\textbf{\base}} & \textbf{TF}  & 4 (2)        & 2 (2)        & 2 (2)        \\
                                               & \textbf{TH}  & 11 (1)       & 1 (0)        & 2 (2)        \\
                                               & \textbf{CK}  & 4 (0)        & 1 (0)        & 0           \\ \bottomrule
            \end{tabular}

        \begin{tablenotes}[flushleft]
        \scriptsize
            \item[1] The number in parentheses denotes the number of voted layers that ONLY detected by the corresponding method.
        \end{tablenotes}

    \end{threeparttable}
    \label{table:voted_layers}
    \vspace{-12pt}
\end{table}

To further illustrate the effectiveness of \rmg, we count the number of distinct voted layers detected by different approaches, as shown in table \ref{table:voted_layers}.  We can observe that all the 4 layers detected by LEMON can be detected by \rmg and \base, indicating that \rmg and \base can cover the exploration scope of LEMON. On the other hand, \rmg and \base have their own distinct voted layers that cannot be detected by the other, proving the effectiveness of inconsistency detection approach. These results indicate that the natural architecture fuzzing approach adopted in \rmg is a good supplement to unit testing.

\begin{table*}[t]
    \centering
    \caption{Comparison of inconsistency number}
    \renewcommand{\arraystretch}{0.9}
    \resizebox{\textwidth}{!}{
        \begin{threeparttable}

        \begin{tabular}{@{}l|l|l|ccc|ccc|ccc|ccc|ccc|ccc|ccc@{}}
            \toprule
            \multirow{2}{*}{\textbf{ID}} & \multirow{2}{*}{\textbf{Method}}   & \multirow{2}{*}{\textbf{Lib Pair}} & \multicolumn{3}{c|}{\textbf{CIFAR-10}}  & \multicolumn{3}{c|}{\textbf{MNIST}}     & \multicolumn{3}{c|}{\textbf{F-MNIST}} & \multicolumn{3}{c|}{\textbf{ImageNet}}  & \multicolumn{3}{c|}{\textbf{Sine-Wave}} & \multicolumn{3}{c|}{\textbf{Stock-Price}} & \multicolumn{3}{c}{\textbf{Total}}     \\
                                            &                                    &                                    & \textbf{FC} & \textbf{LC} & \textbf{BC} & \textbf{FC} & \textbf{LC} & \textbf{BC} & \textbf{FC}   & \textbf{LC}  & \textbf{BC}  & \textbf{FC} & \textbf{LC} & \textbf{BC} & \textbf{FC} & \textbf{LC} & \textbf{BC} & \textbf{FC}  & \textbf{LC}  & \textbf{BC} & \textbf{FC} & \textbf{LC} & \textbf{BC} \\ \midrule
            \multirow{9}{*}{\textbf{E1}}    & \multirow{3}{*}{\textbf{\rmg}}      & \textbf{TF-TH}                     &4             &0             &0             &4             &0             &0             &7               &1              &1              &1             &1             &0             &5              &0             &0             &3              &1              &0             &16             &2             &1             \\
                                            &                                    & \textbf{TF-CK}                     &3             &1             &0             &4             &2             &1             &6               &2              &1              &3             &1             &0             &12             &0             &0             &8              &0              &0             &16             &2             &2             \\
                                            &                                    & \textbf{TH-CK}                     &4             &1             &0             &4             &0             &0             &7               &1              &1              &1             &0             &0             &6              &0             &0             &2              &0              &0             &13             &1             &1             \\ \cmidrule(l){2-24} 
                                            & \multirow{3}{*}{\textbf{LEMON}}    & \textbf{TF-TH}                     &1             &-             &-             &0             &-             &-             &1               &-              &-              &1             &-             &-             &0             &-             &-             &0              &-              &-             &2             &-             &-             \\
                                            &                                    & \textbf{TF-CK}                     &1             &-             &-             &0             &-             &-             &0               &-              &-              &1             &-             &-             &1             &-             &-             &0              &-              &-             &3             &-             &-             \\
                                            &                                    & \textbf{TH-CK}                     &0             &-             &-             &0             &-             &-             &1               &-              &-              &0             &-             &-             &1             &-             &-             &0              &-              &-             &2             &-             &-             \\ \cmidrule(l){2-24} 
                                            & \multirow{3}{*}{\textbf{\base}} & \textbf{TF-TH}                     &1             &0             &3             &4             &2             &2             &2               &0              &2              &5             &0             &0             &1             &0             &0             &0              &1              &0             &9             &2             &3             \\
                                            &                                    & \textbf{TF-CK}                     &2             &1             &1             &1             &2             &1             &1               &3              &1              &2             &2             &0             &2             &0             &0             &4              &1              &0             &5             &3             &1             \\
                                            &                                    & \textbf{TH-CK}                     &1             &0             &0             &3             &1             &1             &2               &0              &1              &5             &0             &0             &2             &0             &0             &4              &0              &0             &10            &1             &1             \\ \bottomrule
        \end{tabular}

        \begin{tablenotes}
            \small
            \item[1] "FC""LC""BC" respectively represent the number of  inconsistencies detected in the three stages. 
            \item[2] For inconsistencies caused by the same kind of layer (or loss function), we only count once.
        \end{tablenotes}

        \end{threeparttable}
    }
  \label{table:incons}
\end{table*}

\begin{table*}[t]
    \centering
    \caption{Comparison of NaN/Crash number}
    \renewcommand{\arraystretch}{0.9}
    \resizebox{\textwidth}{!}{
     \begin{threeparttable}

        \begin{tabular}{@{}l|l|c|ccc|ccc|ccc|ccc|ccc|ccc|ccc@{}}
            \toprule
            \multirow{2}{*}{\textbf{ID}} & \multirow{2}{*}{\textbf{Method}}   & \multirow{2}{*}{\textbf{Lib}} & \multicolumn{3}{c|}{\textbf{CIFAR-10}} & \multicolumn{3}{c|}{\textbf{MNIST}} & \multicolumn{3}{c|}{\textbf{F-MNIST}} & \multicolumn{3}{c|}{\textbf{ImageNet}} & \multicolumn{3}{c|}{\textbf{Sine-Wave}} & \multicolumn{3}{c|}{\textbf{Stock-Price}} & \multicolumn{3}{c}{\textbf{Total}} \\
                                         &                                    &                            & \textbf{NaN}      & \textbf{GC}  & \textbf{EC}     & \textbf{NaN}    & \textbf{GC}  & \textbf{EC}    & \textbf{NaN}        & \textbf{GC}  & \textbf{EC}        & \textbf{NaN}      & \textbf{GC}  & \textbf{EC}     & \textbf{NaN}      & \textbf{GC}  & \textbf{EC}      & \textbf{NaN}       & \textbf{GC}  & \textbf{EC}       & \textbf{NaN}    & \textbf{GC}  & \textbf{EC}    \\ \midrule
            \multirow{9}{*}{\textbf{E1}} & \multirow{3}{*}{\textbf{\rmg}}      & \textbf{TF}           &2                   &0  &1                 &3                 &0 &1                 &3                     &0 &1                     &5                   &0 &1                  &4                   &0 &1                   &5                    &0 &1                    &7                 &0 &1                    \\
                                         &                                    & \textbf{TH}           &3                   &0  &10                &2                 &0 &10                &1                     &0 &3                     &2                   &0 &3                  &3                   &0 &8                   &3                    &0 &4                    &6                 &0 &10                   \\
                                         &                                    & \textbf{CK}           &2                   &6  &4                 &3                 &6 &4                 &3                     &4 &3                     &4                   &2 &2                  &4                   &4 &4                   &5                    &4 &4                    &7                 &6 &4                    \\ \cmidrule(l){2-24} 
                                         & \textbf{LEMON}                     & \textbf{-}            &0                   &-  &0                 &0                 &-  &0                &0                     &-  &0                    &0                   &-  &0                 &0                   &-  &0                  &0                    &-  &0                   &0                 &-  &0                \\ \cmidrule(l){2-24} 
                                         & \multirow{3}{*}{\textbf{\base}} & \textbf{TF}           &0                   &0  &0                 &0                 &0  &0                &0                     &0  &0                    &0                   &0  &0                 &4                   &0  &0                  &0                    &0  &0                   &4                 &0  &0                \\
                                         &                                    & \textbf{TH}           &0                   &0  &3                 &0                 &0  &3                &0                     &0  &2                    &0                   &0  &2                 &4                   &0  &3                  &0                    &0  &3                   &4                 &0  &3                \\
                                         &                                    & \textbf{CK}           &0                   &1  &4                 &0                 &0  &4                &0                     &0  &3                    &0                   &0  &2                 &5                   &0  &4                  &0                    &0  &3                   &5                 &1  &4                \\ \bottomrule

            \end{tabular}

        \begin{tablenotes}
            \small
            \item[1] "NaN" represents the number of outputs with NaN. For NaN caused by the same kind of layer, we only count once.
            \item[2] "GC" and "EC" are respectively short for "Generation Crash" and "Execution Crash". The crash number have been deduplicated according to error messages.
            \item[3] LEMON does not record NaN/crash information for each backend, so we only obtain the total NaN/crash number triggered by LEMON.
        \end{tablenotes}

    \end{threeparttable}
    }
    \label{table:nan_crash}
\end{table*}

\subsection{Performance Comparison}

In order to further evaluate the performance of \rmg, we compare the number of inconsistencies, NaN, and crash detected by different methods. We present the results under environment E1 as an example\footnote{More experiment codes and results are available at https://github.com/library-testing/Muffin}. Table \ref{table:incons} shows the inconsistencies detected by three methods under different datasets and environments. Specifically, in the latest library versions (\ie E1), \rmg finds a total of 54 inconsistencies, 45 of which are found in the FC stage. In comparison, LEMON can only find 7 inconsistencies, much less than \rmg. Similar results can also be observed in other environments, which prove the effectiveness of \rmg in library testing. The main reason is that \rmg can explore more library functions through the model generation approach, while LEMON can only mutate seed models, and thus can hardly cover the functions not being used in seed models. Besides, LEMON also cannot explore the library codes related to loss and gradient calculation. As a result, LEMON only achieves 35.593\% functionality coverage (the percentage of the invoked APIs in all the pre-defined, learning-related APIs we considered), while \rmg can achieve 98.305\% functionality coverage. The inconsistent APIs that cannot be identified by LEMON include ``DepthwiseConv2D'', ``LocallyConnected1D'', ``Conv3D'' and various loss functions. It is worth noting that although \rmg is not designed to achieve high line coverage, we summarize and report the line coverage results: \rmg achieves 43.22\%, which is 2.07 times of that achieved by LEMON (20.85\%). 

On the other hand, compared with \base, \rmg detects 19 more inconsistencies in E1, which proves the performance of \rmg. It is also worth noting that the number of inconsistencies detected by \rmg reduces in other environments. The key reason is that the numbers of NaN and crash triggered by \rmg increase in old library versions, as shown in Table \ref{table:nan_crash}. Taking NaN and crash into consideration, \rmg can still trigger more exceptions (\ie inconsistency, NaN and crash) than \base. In particular, the layer functions where \rmg can detect exceptions while \base cannot include ``AveragePooling1D'', ``Conv3DTranspose'' and ``CategoricalCrossentropy''.



\begin{table}[t!]
    \centering
    \caption{Comparison of execution time (MIN.)}
    \renewcommand{\arraystretch}{0.9}
    \resizebox{0.55\linewidth}{!}{
     \begin{threeparttable}

        \begin{tabular}{@{}l|l|ccccc@{}}
            \toprule
            \textbf{Dataset}                        & \textbf{Method} & \textbf{E1} & \textbf{E2} & \textbf{E3} & \textbf{E4} & \textbf{E5} \\ \midrule
            \multirow{3}{*}{\textbf{CIFAR-10}}      & \rmg             & 29.20       & 20.67       & 38.75       & 27.38       & 16.75       \\
                                                    & LEMON           & 27.20       & 28.02       & 32.48       & 34.70       & 28.60       \\
                                                    & \base        & 16.27       & 14.72       & 13.13       & 11.00       & 13.47       \\ \midrule
            \multirow{3}{*}{\textbf{MNIST}}         & \rmg             & 32.35       & 19.82       & 18.20       & 14.95       & 15.78       \\
                                                    & LEMON           & 10.58       & 10.28       & 9.88        & 9.67        & 9.28        \\
                                                    & \base        & 14.87       & 14.88       & 12.40       & 11.02       & 13.20       \\ \midrule
            \multirow{3}{*}{\textbf{F-MNIST}} & \rmg             & 24.78       & 25.00       & 18.23       & 15.92       & 20.38       \\
                                                    & LEMON           & 12.62       & 12.88       & 12.27       & 11.67       & 11.27       \\
                                                    & \base        & 15.85       & 15.32       & 12.90       & 11.67       & 12.78       \\ \midrule
            \multirow{3}{*}{\textbf{ImageNet}}      & \rmg             & 49.04       & 40.72       & 38.12       & 34.50       & 28.22       \\
                                                    & LEMON           & 80.25       & 117.62      & 114.25      & 117.25      & 111.93      \\
                                                    & \base        & 34.03       & 23.52       & 30.15       & 37.05       & 27.55       \\ \midrule
            \multirow{3}{*}{\textbf{Sine-Wave}}     & \rmg             & 25.95       & 24.37       & 18.52       & 15.40       & 17.45       \\
                                                    & LEMON           & 15.37       & 14.37       & 13.67       & 13.37       & 13.01       \\
                                                    & \base        & 17.78       & 16.45       & 13.40       & 12.60       & 14.83       \\ \midrule
            \multirow{3}{*}{\textbf{Stock-Price}}   & \rmg             & 22.03       & 18.78       & 16.35       & 13.28       & 14.28       \\
                                                    & LEMON           & 16.58       & 15.50       & 14.35       & 14.37       & 13.82       \\
                                                    & \base        & 16.07       & 14.22       & 12.45       & 10.85       & 13.25       \\ \bottomrule
            \end{tabular}


    \end{threeparttable}
    }
    \label{table:execution_time}
    \vspace{-10pt}
\end{table}

Furthermore, we also compare the execution time of the three methods to generate 50 models and perform testing under different datasets. The execution time of \rmg and \base consists of the model generation time and the three-stage inconsistency detection time (\ie FC, LC and BC). The execution time of LEMON consists of model mutation time and inconsistency detection time (only FC).
The results are shown in Table \ref{table:execution_time}.

In this table, we can observe that except ImageNet, the execution time of \rmg is the longest in most cases. The reason is that \rmg conducts additional model generation (compared with \base), and inconsistency detection in additional two stages (compare with LEMON). Considering that \rmg can detect much more inconsistencies, we think such overhead (\ie around ten minutes) is acceptable. These results also demonstrate that the proposed approach (model generation and inconsistency detection) do not bring huge overhead to \rmg.

Moreover, when performing library testing with ImageNet, the execution time of LEMON is greatly increased. The reason is that the seed models used by LEMON are much more complicated, compared to those under other datasets. This phenomenon reveals that the execution time of LEMON highly depends on the complexity of seed models. On the other hand, \rmg does not suffer from this problem. The generated model complexity of \rmg can be controlled via setting proper values of $MAX_c$ and $MAX_v$. Under the same $MAX_c$ and $MAX_v$ value, the execution time of \rmg is quite stable.

Compared with \base, \rmg requires additional DAG-based model generation. In addition, the number of layers in the model also affects the inconsistency detection time. The larger the model, the longer the detection time. As a result, the execution time of \rmg is slightly longer than that of \base.

Finally, it is worth noting \rmg does not consider the final model performance (\eg precision and recall) in specific tasks when generating model architectures, since it is not the objective of a testing tool. In contrast, existing approaches (\eg mutating existing models) typically generate limited model architectures but can obtain high-performance models, which however, are not more capable in detecting bugs.
Instead, \rmg focuses more on model quality in testing. \rmg can generate high-quality models. In 900 executions (3 libraries, each with 300 models), only 77 executions (8.5\%) cause four unsupported-crashes (by the same reason). Moreover, we have also shown such models are more capable in exposing bugs, as discussed in Section \ref{sec:results:rq1}.

\subsection{Effect of Different Parameter Settings}
\label{sec:results:rq3}
\rmg introduces four parameters, \ie $MAX_c$ and $MAX_v$ to control the size of model structure, and threshold $t$ and $\epsilon$ for inconsistency detection. Since in all experiments, \rmg achieves satisfying layer coverage (\ie only one layer cannot be used with all datasets), we consider that the values of $MAX_c$ and $MAX_v$ are set properly. For the thresholds, $\epsilon$ is a quite small value (\ie $1e^{-5}$), thus we only evaluate the number of inconsistencies detected by \rmg with different $t$ values.

\pgfkeys{/pgfplots/legend entry/.code=\addlegendentry{#1}}

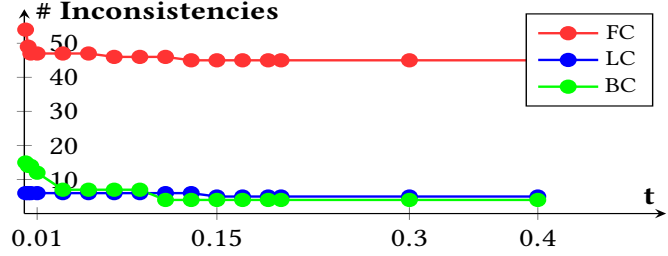
\begin{figure}[t!]
    \centering
    \resizebox{0.6\linewidth}{3.5cm}{
        \begin{tikzpicture}[
            every mark/.append style = {solid},
        ]
            \begin{axis}[
                xmin = 0,
                xmax = 0.5,
                xlabel = t,
                xtick = {0.01, 0.15, 0.3, 0.4},
                xticklabel style={
                /pgf/number format/fixed,
                },
                axis x line=middle,
                ymax= 60,
                ymin = 0,
                ytick={0, 10, 20, 30, 40, 50},
                width = 8cm,
                height = 4cm,
                axis y line = middle,
                ylabel={\# Inconsistencies},
                ylabel style = {yshift=6pt, font=\bfseries},
                xlabel style = {font=\bfseries},
                yticklabel style = {right, xshift=6pt},
                table/col sep=semicolon,
                legend style={nodes={scale=0.8, transform shape}}, 
            ]
                \addplot[mark=*, color=red!80, thick] table[x=Threshold, y=Value, legend entry=FC] {FCIncons.csv};
                
                \addplot[mark=*, color=blue, thick] table[x=Threshold, y=Value, legend entry=LC] {LCIncons.csv};
                
                \addplot[mark=*, color=green, thick] table[x=Threshold, y=Value, legend entry=BC] {BCIncons.csv};
            \end{axis}
            
        \end{tikzpicture}
        }
    \caption{Performance of \rmg under different thresholds}
    \label{fig:threshold}
    \vspace{-12pt}
\end{figure}

Figure~\ref{fig:threshold} shows the number of inconsistencies detected by \rmg under different $t$ values, ranging from $0.001$ to $0.4$. We can observe that when $t$ is small, \rmg is more sensitive to small variance of differences, thus it detects more inconsistencies. As the value of $t$ increases, the number of inconsistencies decreases slowly, and keeps stable when $t \in [0.15, 0.4]$. Thus, the default $t$ value in \rmg is $0.15$. Although bugs incurring small variances may be neglected, such bugs can be revealed under other input values or model architectures (\ie variance larger than $t$).
\section{Discussion}
\label{sec:discussion}
\subsection{Summary of Evaluation}
As discussed before, \base is designed based on the idea of unit testing, which tests a specific library function at a time. The evaluation results show that \rmg can detect more layer inconsistencies than \base. The main reason is that many layer inconsistencies can only be triggered by specific inputs. For instance, the gradients inconsistency of ``MaxPooling1D'' layer only happens when multiple elements in the input tensor have the same maximum value. In order to trigger such inconsistencies using \base, we have to fuzzing the inputs. Since layers in DL libraries typically have huge input value ranges, \eg high-dimensional tensor inputs where each element ranges from $(-\infty, \infty)$, it is quite challenging to find specific inputs that can trigger corner cases \cite{DBLP:conf/dsn/WuXZLK19}. On the other hand, \rmg performs testing based on generated models. Due to the existence of different layer types, we thus simulates the real calculation process and reduce the input ranges. As a result, the possible corner cases (\ie multiple maximum values) can be triggered by \rmg. Considering that unit testing is necessary before version release, while bugs can still be detected in the latest versions, we believe DL library testing based on model generation is an effective supplement to unit testing.

Among the bugs detected by \rmg, some of them are actually caused by unclear specifications. For instance, the gradient calculation bug of ``categorical\_hinge'' loss function is actually caused by the different specification of calculating the gradients of ``max()'' function. Specifically, when there are multiple maximum elements, TensorFlow will divide the gradient with the number of maximum element, while Theano and CNTK do not have this operation. Similarly, for ``MaxPooling1D'' layer, when there are multiple maximum elements, TensorFlow and CNTK would only apply the gradients to one of the maximum elements, while Theano apply the gradients to all maximum elements. Due to unclear specifications, different DL libraries have different implementations. Although in most cases the results of these implementations are consistent, robustness issues may be caused by the corner cases (\eg easier to generate adversarial inputs \cite{DBLP:conf/nips/GoodfellowPMXWOCB14}). Therefore, we call for the community to pay more attention on the unclear specification problems in DL libraries.

\subsection{Threats to Validity}
We now discuss possible threats in this work, and the methods we take to address such threats. 
First of all, we only evaluate the effectiveness of \rmg on three DL libraries, \ie TensorFlow, Theano and CNTK. These libraries can be called using the same front-end library (\ie Keras), which facilitate the implementation and performing differential testing. Other libraries that do not support Keras (\ie PyTorch) currently are not supported by \rmg. However, the ideas of model generation and inconsistency detection adopted in \rmg are general. For instance, in order to test PyTorch, it requires to replace the Keras APIs used in \rmg with the corresponding PyTorch APIs. To reduce this treat, we evaluate \rmg with a total of 15 different release versions of DL libraries. In addition, we also use diverse models (including the models generated by \rmg, existing models on 6 real datasets, as well as their mutants generated by existing work) to evaluate the inconsistency detection performance of our approach.

Another threat mainly lies in randomness and threshold settings in our experiment. To reduce the randomness, we conduct five experiments with different library versions (\ie E1-E5, refer to Table 1, Table2 in Supplementary Material). In each experiment, every method generates/mutates the same number of models for 6 commonly-used datasets, and we record and compare the results and execution time. For threshold settings (\eg $t$, $MAX_c$ and $MAX_v$), since in every experiment \rmg achieves 58/59 function usage, we do not increase $MAX_c$ and $MAX_v$. For threshold $t$, as discussed in Section \ref{sec:evaluation:imp}, we choose a quite large threshold. Slightly changing $t$ (\eg from 0.15 to 0.4) has little impact on the results.

\subsection{Future Directions}
\rmg can be potentially improved in the following two aspects.
First, \rmg only covers library codes in the layer function granularity through trying to generate model covering all the provided APIs. However, there may still be a large portion of library codes that cannot be covered (\eg private methods, branches). In the future, \rmg can be extended to consider other coverage metrics (\eg line coverage, branch coverage), and conducts model generation/mutation to explore more library codes.

Second, \rmg still relies on differential testing to solve the test oracle problem. However, if different DL libraries produce the same wrong results, \rmg cannot identify such bugs. Moreover, in the evaluation, we also notice that under certain circumstances, the model generated by \rmg may cause one library to crash, while the other two produce inconsistent results. In such cases, it requires huge human efforts to identify potential bugs. To get rid of this limitation, we intend to design metamorphic relations based on the properties of DL models, and conducts metamorphic testing to test one library accordingly.
\section{Related Work}
\label{sec:related}
As discussed before, CRADLE~\cite{DBLP:conf/icse/PhamLQT19} and LEMON~\cite{DBLP:conf/sigsoft/WangYCLZ20} are the most related work to ours that targets DL library testing, both of which require existing DL models and only detect bugs in model inference phase. Different from them, \rmg detects DL library bugs in model training phase via DAG-based model generation. In the literature, there is a body of work focusing on testing machine learning (ML) libraries as well~\cite{zhang2020machine, DBLP:journals/tr/XieZCLPX20, DBLP:conf/sigsoft/DuttaLHM18, DBLP:conf/sigsoft/0001ZHM19, DBLP:conf/issta/DwarakanathASRB18}. For instance, Dutta \etal~\cite{DBLP:conf/sigsoft/DuttaLHM18} propose ProbFuzz to test probabilistic programming systems via generating programs based on pre-defined templates. Dwarakanath \etal~\cite{DBLP:conf/issta/DwarakanathASRB18} adopt metamorphic testing to test image classification applications through mutating the training and testing data. However, these approaches cannot be directly adopted for DL libraries testing.

On the other hand, there is a great deal of researches focusing on the testing of DL models~\cite{DBLP:conf/sosp/PeiCYJ17, DBLP:conf/icse/TianPJR18, DBLP:conf/kbse/MaJZSXLCSLLZW18, DBLP:conf/issre/MaZSXLJXLLZW18, DBLP:conf/sigsoft/MaLLZG18, DBLP:conf/dsn/WuXZLK19, DBLP:conf/icse/KimFY19, DBLP:conf/issta/XieMJXCLZLYS19, DBLP:conf/icml/OdenaOAG19}. In particular, many research efforts have been put on designing criteria to measure test adequacy~\cite{DBLP:conf/issre/MaZSXLJXLLZW18, DBLP:conf/icse/KimFY19, DBLP:conf/sigsoft/DuXLM0Z19}. For instance, Pei \etal~\cite{DBLP:conf/sosp/PeiCYJ17} first propose neuron coverage as the criteria for testing DL models. Ma \etal~\cite{DBLP:conf/issre/MaZSXLJXLLZW18} further define both neuron and layer level coverage criteria to help gauging the testing quality of DL models. Kim \etal~\cite{DBLP:conf/icse/KimFY19} propose surprise coverage based on surprise adequacy, which measures relative surprise of each input with respect to the training data. Du \etal~\cite{DBLP:conf/sigsoft/DuXLM0Z19} propose a set of similarity metrics and coverage criteria to analyze stateful DL systems such as Recurrent Neural Networks (RNNs)~\cite{DBLP:conf/interspeech/MikolovKBCK10}.
Moreover, there are a lot of studies intend to reveal defects in DL models via generating adversarial inputs or finding corner cases~\cite{DBLP:conf/icse/TianPJR18, DBLP:conf/kbse/ZhangZZ0K18, DBLP:conf/issta/XieMJXCLZLYS19}. For instance, Tian \etal~\cite{DBLP:conf/icse/TianPJR18} implement DeepTest for detecting erroneous behaviors of DL-based self-driving cars via automatically generating test cases based on image transformations. Similarly, Zhang \etal~\cite{DBLP:conf/kbse/ZhangZZ0K18} implement DeepRoad, which applies Generative Adversarial Networks (GANs)~\cite{DBLP:conf/nips/GoodfellowPMXWOCB14} to test DL-based self-driving cars. 
Besides, there are also many researches focus on detecting different kinds of bugs in model structures or training parameter settings~\cite{DBLP:conf/sigsoft/ZhangRC0C020, DBLP:conf/icse/WardatLR21}. For instance, Zhang \etal~\cite{DBLP:conf/sigsoft/ZhangRC0C020} propose DEBAR, a static analysis approach for detecting numerical bugs in DL models. Wardat \etal~\cite{DBLP:conf/icse/WardatLR21} propose a dynamic analysis based approach to detect numerical errors when training DL models. Similarly, Zhang \etal~\cite{DBLP:conf/icse/ZhangZMS21} propose AUTOTRAINER, a tool that detects and auto-repairs commonly-seen model training problems such as vanishing gradient, exploding gradients and slow convergence. Different from them, our work focuses on testing DL libraries rather than DL models or parameters.

Our work is also related to differential testing, an effective method that use similar programs as cross referencing oracles to detect bugs~\cite{DBLP:conf/icse/GulzarZH19}. Differential testing has been successful in uncovering bugs across various types of programs, such as compilers~\cite{DBLP:conf/pldi/YangCER11}, Java Virtual Machine (JVM) implementations~\cite{DBLP:conf/pldi/ChenSSSZ16, DBLP:conf/icse/ChenSS19}, web applications~\cite{DBLP:conf/ccs/ChapmanE11}, and security-related APIs~\cite{DBLP:conf/pldi/SrivastavaBMS11}. In recent years, researchers also utilize differential testing in the area of DL testing~\cite{DBLP:conf/sosp/PeiCYJ17, DBLP:conf/icse/TianPJR18}. For instance, Pei \etal~\cite{DBLP:conf/sosp/PeiCYJ17} propose DeepXplore, a differential testing framework to identify DL model defects via image transformation. Guo \etal~\cite{DBLP:conf/sigsoft/GuoJZCS18} propose DLFuzz, a differential fuzzing testing framework that exposes DL model errors through mutating inputs to maximize model output difference. These approaches focus on testing DL models, while \rmg is designed for DL library testing with high coverage.

\section{Conclusion}
\label{sec:conclusion}
In this paper, we propose a novel approach to test DL library codes via direct model generation using library APIs. In order to generate diverse DL models, we use DAG to formulate the model structure and propose a DAG-based model generation algorithm. In order to detect bugs, we divide the model training phase into three stages, and design different measurements for each stage. In this way, our approach can detect library bugs related to model training, which is not covered by previous studies. We implement our approach as an open-source tool called \rmg. To evaluate the performance of \rmg, we conduct a series of experiments based on 15 release versions of three widely-used DL libraries. \rmg detects 39 new bugs in the latest versions of these libraries. Besides, \rmg outperforms other methods in terms of the number of detected unique inconsistencies.

\begin{acks}
This work was supported by the National Key R\&D Program of China under Grant 2020YFA0711400 and the Natural Science Foundation of Shanghai (No. 22ZR1407900).
\end{acks}

\balance
\bibliographystyle{ACM-Reference-Format}
\bibliography{dl_lib_testing}


\begin{thebibliography}{62}


\ifx \showCODEN    \undefined \def \showCODEN     #1{\unskip}     \fi
\ifx \showDOI      \undefined \def \showDOI       #1{#1}\fi
\ifx \showISBNx    \undefined \def \showISBNx     #1{\unskip}     \fi
\ifx \showISBNxiii \undefined \def \showISBNxiii  #1{\unskip}     \fi
\ifx \showISSN     \undefined \def \showISSN      #1{\unskip}     \fi
\ifx \showLCCN     \undefined \def \showLCCN      #1{\unskip}     \fi
\ifx \shownote     \undefined \def \shownote      #1{#1}          \fi
\ifx \showarticletitle \undefined \def \showarticletitle #1{#1}   \fi
\ifx \showURL      \undefined \def \showURL       {\relax}        \fi
\providecommand\bibfield[2]{#2}
\providecommand\bibinfo[2]{#2}
\providecommand\natexlab[1]{#1}
\providecommand\showeprint[2][]{arXiv:#2}

\bibitem[\protect\citeauthoryear{??}{url}{2021a}]%
        {url:TPU}
 \bibinfo{year}{Accessed: 2021}\natexlab{a}.
\newblock \bibinfo{title}{Cloud TPU}.
\newblock \bibinfo{howpublished}{\url{https://cloud.google.com/tpu}}.
\newblock


\bibitem[\protect\citeauthoryear{??}{url}{2021b}]%
        {url:CNTK}
 \bibinfo{year}{Accessed: 2021}\natexlab{b}.
\newblock \bibinfo{title}{CNTK}.
\newblock
  \bibinfo{howpublished}{\url{https://docs.microsoft.com/cognitive-toolkit}}.
\newblock


\bibitem[\protect\citeauthoryear{??}{url}{2021c}]%
        {url:CNTK_ops}
 \bibinfo{year}{Accessed: 2021}\natexlab{c}.
\newblock \bibinfo{title}{CNTK ops Pachakge: sqrt}.
\newblock
  \bibinfo{howpublished}{\url{https://docs.microsoft.com/en-us/python/api/cntk/cntk.ops?view=cntk-py-2.7##sqrt-x--name----}}.
\newblock


\bibitem[\protect\citeauthoryear{??}{url}{2021d}]%
        {url:Keras}
 \bibinfo{year}{Accessed: 2021}\natexlab{d}.
\newblock \bibinfo{title}{Keras}.
\newblock \bibinfo{howpublished}{\url{https://keras.io}}.
\newblock


\bibitem[\protect\citeauthoryear{??}{url}{2021e}]%
        {url:TensorFlow}
 \bibinfo{year}{Accessed: 2021}\natexlab{e}.
\newblock \bibinfo{title}{TensorFlow}.
\newblock \bibinfo{howpublished}{\url{https://www.tensorflow.org}}.
\newblock


\bibitem[\protect\citeauthoryear{??}{url}{2021f}]%
        {url:tesla2016}
 \bibinfo{year}{Accessed: 2021}\natexlab{f}.
\newblock \bibinfo{title}{Tesla driver dies in first fatal crash while using
  autopilot mode}.
\newblock
  \bibinfo{howpublished}{\url{https://www.theguardian.com/technology/2016/jun/30/tesla-autopilot-death-self-driving-car-elon-musk}}.
\newblock


\bibitem[\protect\citeauthoryear{??}{url}{2021g}]%
        {url:uber2018}
 \bibinfo{year}{Accessed: 2021}\natexlab{g}.
\newblock \bibinfo{title}{Uber's self-driving operator charged over fatal
  crash}.
\newblock
  \bibinfo{howpublished}{\url{https://www.bbc.com/news/technology-54175359}}.
\newblock


\bibitem[\protect\citeauthoryear{Abadi, Barham, Chen, Chen, Davis, Dean, Devin,
  Ghemawat, Irving, Isard, Kudlur, Levenberg, Monga, Moore, Murray, Steiner,
  Tucker, Vasudevan, Warden, Wicke, Yu, and Zheng}{Abadi et~al\mbox{.}}{2016}]%
        {DBLP:conf/osdi/AbadiBCCDDDGIIK16}
\bibfield{author}{\bibinfo{person}{Mart{\'{\i}}n Abadi}, \bibinfo{person}{Paul
  Barham}, \bibinfo{person}{Jianmin Chen}, \bibinfo{person}{Zhifeng Chen},
  \bibinfo{person}{Andy Davis}, \bibinfo{person}{Jeffrey Dean},
  \bibinfo{person}{Matthieu Devin}, \bibinfo{person}{Sanjay Ghemawat},
  \bibinfo{person}{Geoffrey Irving}, \bibinfo{person}{Michael Isard},
  \bibinfo{person}{Manjunath Kudlur}, \bibinfo{person}{Josh Levenberg},
  \bibinfo{person}{Rajat Monga}, \bibinfo{person}{Sherry Moore},
  \bibinfo{person}{Derek~Gordon Murray}, \bibinfo{person}{Benoit Steiner},
  \bibinfo{person}{Paul~A. Tucker}, \bibinfo{person}{Vijay Vasudevan},
  \bibinfo{person}{Pete Warden}, \bibinfo{person}{Martin Wicke},
  \bibinfo{person}{Yuan Yu}, {and} \bibinfo{person}{Xiaoqiang Zheng}.}
  \bibinfo{year}{2016}\natexlab{}.
\newblock \showarticletitle{TensorFlow: {A} System for Large-Scale Machine
  Learning}. In \bibinfo{booktitle}{\emph{Proc. of the 12th {USENIX} Symposium
  on Operating Systems Design and Implementation, {OSDI}}}.
  \bibinfo{publisher}{{USENIX} Association}, \bibinfo{pages}{265--283}.
\newblock


\bibitem[\protect\citeauthoryear{Cantrell}{Cantrell}{2000}]%
        {cantrell2000modern}
\bibfield{author}{\bibinfo{person}{Cyrus~D Cantrell}.}
  \bibinfo{year}{2000}\natexlab{}.
\newblock \bibinfo{booktitle}{\emph{Modern mathematical methods for physicists
  and engineers}}.
\newblock \bibinfo{publisher}{Cambridge University Press}.
\newblock


\bibitem[\protect\citeauthoryear{Chapman and Evans}{Chapman and Evans}{2011}]%
        {DBLP:conf/ccs/ChapmanE11}
\bibfield{author}{\bibinfo{person}{Peter Chapman} {and} \bibinfo{person}{David
  Evans}.} \bibinfo{year}{2011}\natexlab{}.
\newblock \showarticletitle{Automated black-box detection of side-channel
  vulnerabilities in web applications}. In \bibinfo{booktitle}{\emph{Proc. of
  the 18th {ACM} Conference on Computer and Communications Security, {CCS}}}.
  \bibinfo{publisher}{{ACM}}, \bibinfo{pages}{263--274}.
\newblock
\urldef\tempurl%
\url{https://doi.org/10.1145/2046707.2046737}
\showDOI{\tempurl}


\bibitem[\protect\citeauthoryear{Chen, He, Lin, Zhang, Hao, Gao, Xu, Dang, and
  Zhang}{Chen et~al\mbox{.}}{2019a}]%
        {DBLP:conf/kbse/0003HL0HGXDZ19}
\bibfield{author}{\bibinfo{person}{Junjie Chen}, \bibinfo{person}{Xiaoting He},
  \bibinfo{person}{Qingwei Lin}, \bibinfo{person}{Hongyu Zhang},
  \bibinfo{person}{Dan Hao}, \bibinfo{person}{Feng Gao},
  \bibinfo{person}{Zhangwei Xu}, \bibinfo{person}{Yingnong Dang}, {and}
  \bibinfo{person}{Dongmei Zhang}.} \bibinfo{year}{2019}\natexlab{a}.
\newblock \showarticletitle{Continuous Incident Triage for Large-Scale Online
  Service Systems}. In \bibinfo{booktitle}{\emph{Proc. of the 34th {IEEE/ACM}
  International Conference on Automated Software Engineering, {ASE}}}.
  \bibinfo{publisher}{{IEEE}}, \bibinfo{pages}{364--375}.
\newblock
\urldef\tempurl%
\url{https://doi.org/10.1109/ASE.2019.00042}
\showDOI{\tempurl}


\bibitem[\protect\citeauthoryear{Chen, Zhang, He, Lin, Zhang, Hao, Kang, Gao,
  Xu, Dang, and Zhang}{Chen et~al\mbox{.}}{2020}]%
        {DBLP:conf/kbse/ChenZHLZHKGXDZ20}
\bibfield{author}{\bibinfo{person}{Junjie Chen}, \bibinfo{person}{Shu Zhang},
  \bibinfo{person}{Xiaoting He}, \bibinfo{person}{Qingwei Lin},
  \bibinfo{person}{Hongyu Zhang}, \bibinfo{person}{Dan Hao},
  \bibinfo{person}{Yu Kang}, \bibinfo{person}{Feng Gao},
  \bibinfo{person}{Zhangwei Xu}, \bibinfo{person}{Yingnong Dang}, {and}
  \bibinfo{person}{Dongmei Zhang}.} \bibinfo{year}{2020}\natexlab{}.
\newblock \showarticletitle{How Incidental are the Incidents? Characterizing
  and Prioritizing Incidents for Large-Scale Online Service Systems}. In
  \bibinfo{booktitle}{\emph{Proc. of the 35th {IEEE/ACM} International
  Conference on Automated Software Engineering, {ASE}}}.
  \bibinfo{publisher}{{IEEE}}, \bibinfo{pages}{373--384}.
\newblock
\urldef\tempurl%
\url{https://doi.org/10.1145/3324884.3416624}
\showDOI{\tempurl}


\bibitem[\protect\citeauthoryear{Chen, Su, and Su}{Chen et~al\mbox{.}}{2019b}]%
        {DBLP:conf/icse/ChenSS19}
\bibfield{author}{\bibinfo{person}{Yuting Chen}, \bibinfo{person}{Ting Su},
  {and} \bibinfo{person}{Zhendong Su}.} \bibinfo{year}{2019}\natexlab{b}.
\newblock \showarticletitle{Deep differential testing of {JVM}
  implementations}. In \bibinfo{booktitle}{\emph{Proc. of the 41st
  International Conference on Software Engineering, {ICSE}}}.
  \bibinfo{publisher}{{IEEE} / {ACM}}, \bibinfo{pages}{1257--1268}.
\newblock
\urldef\tempurl%
\url{https://doi.org/10.1109/ICSE.2019.00127}
\showDOI{\tempurl}


\bibitem[\protect\citeauthoryear{Chen, Su, Sun, Su, and Zhao}{Chen
  et~al\mbox{.}}{2016}]%
        {DBLP:conf/pldi/ChenSSSZ16}
\bibfield{author}{\bibinfo{person}{Yuting Chen}, \bibinfo{person}{Ting Su},
  \bibinfo{person}{Chengnian Sun}, \bibinfo{person}{Zhendong Su}, {and}
  \bibinfo{person}{Jianjun Zhao}.} \bibinfo{year}{2016}\natexlab{}.
\newblock \showarticletitle{Coverage-directed differential testing of {JVM}
  implementations}. In \bibinfo{booktitle}{\emph{Proc. of the 37th {ACM}
  {SIGPLAN} Conference on Programming Language Design and Implementation,
  {PLDI}}}. \bibinfo{publisher}{{ACM}}, \bibinfo{pages}{85--99}.
\newblock
\urldef\tempurl%
\url{https://doi.org/10.1145/2908080.2908095}
\showDOI{\tempurl}


\bibitem[\protect\citeauthoryear{Choi, Fazekas, Sandler, and Cho}{Choi
  et~al\mbox{.}}{2017}]%
        {DBLP:conf/icassp/ChoiFSC17}
\bibfield{author}{\bibinfo{person}{Keunwoo Choi}, \bibinfo{person}{Gy{\"{o}}rgy
  Fazekas}, \bibinfo{person}{Mark~B. Sandler}, {and} \bibinfo{person}{Kyunghyun
  Cho}.} \bibinfo{year}{2017}\natexlab{}.
\newblock \showarticletitle{Convolutional recurrent neural networks for music
  classification}. In \bibinfo{booktitle}{\emph{2017 {IEEE} International
  Conference on Acoustics, Speech and Signal Processing, {ICASSP}}}.
  \bibinfo{publisher}{{IEEE}}, \bibinfo{pages}{2392--2396}.
\newblock
\urldef\tempurl%
\url{https://doi.org/10.1109/ICASSP.2017.7952585}
\showDOI{\tempurl}


\bibitem[\protect\citeauthoryear{Du, Xie, Li, Ma, Liu, and Zhao}{Du
  et~al\mbox{.}}{2019}]%
        {DBLP:conf/sigsoft/DuXLM0Z19}
\bibfield{author}{\bibinfo{person}{Xiaoning Du}, \bibinfo{person}{Xiaofei Xie},
  \bibinfo{person}{Yi Li}, \bibinfo{person}{Lei Ma}, \bibinfo{person}{Yang
  Liu}, {and} \bibinfo{person}{Jianjun Zhao}.} \bibinfo{year}{2019}\natexlab{}.
\newblock \showarticletitle{DeepStellar: model-based quantitative analysis of
  stateful deep learning systems}. In \bibinfo{booktitle}{\emph{Proc. of the
  {ACM} Joint Meeting on European Software Engineering Conference and Symposium
  on the Foundations of Software Engineering, {ESEC/FSE}}}.
  \bibinfo{publisher}{{ACM}}, \bibinfo{pages}{477--487}.
\newblock
\urldef\tempurl%
\url{https://doi.org/10.1145/3338906.3338954}
\showDOI{\tempurl}


\bibitem[\protect\citeauthoryear{Dutta, Legunsen, Huang, and Misailovic}{Dutta
  et~al\mbox{.}}{2018}]%
        {DBLP:conf/sigsoft/DuttaLHM18}
\bibfield{author}{\bibinfo{person}{Saikat Dutta}, \bibinfo{person}{Owolabi
  Legunsen}, \bibinfo{person}{Zixin Huang}, {and} \bibinfo{person}{Sasa
  Misailovic}.} \bibinfo{year}{2018}\natexlab{}.
\newblock \showarticletitle{Testing probabilistic programming systems}. In
  \bibinfo{booktitle}{\emph{Proc. of the 2018 {ACM} Joint Meeting on European
  Software Engineering Conference and Symposium on the Foundations of Software
  Engineering, {ESEC/FSE}}}. \bibinfo{publisher}{{ACM}},
  \bibinfo{pages}{574--586}.
\newblock
\urldef\tempurl%
\url{https://doi.org/10.1145/3236024.3236057}
\showDOI{\tempurl}


\bibitem[\protect\citeauthoryear{Dutta, Zhang, Huang, and Misailovic}{Dutta
  et~al\mbox{.}}{2019}]%
        {DBLP:conf/sigsoft/0001ZHM19}
\bibfield{author}{\bibinfo{person}{Saikat Dutta}, \bibinfo{person}{Wenxian
  Zhang}, \bibinfo{person}{Zixin Huang}, {and} \bibinfo{person}{Sasa
  Misailovic}.} \bibinfo{year}{2019}\natexlab{}.
\newblock \showarticletitle{Storm: program reduction for testing and debugging
  probabilistic programming systems}. In \bibinfo{booktitle}{\emph{Proc. of the
  {ACM} Joint Meeting on European Software Engineering Conference and Symposium
  on the Foundations of Software Engineering, {ESEC/FSE}}}.
  \bibinfo{publisher}{{ACM}}, \bibinfo{pages}{729--739}.
\newblock
\urldef\tempurl%
\url{https://doi.org/10.1145/3338906.3338972}
\showDOI{\tempurl}


\bibitem[\protect\citeauthoryear{Dwarakanath, Ahuja, Sikand, Rao, Bose, Dubash,
  and Podder}{Dwarakanath et~al\mbox{.}}{2018}]%
        {DBLP:conf/issta/DwarakanathASRB18}
\bibfield{author}{\bibinfo{person}{Anurag Dwarakanath}, \bibinfo{person}{Manish
  Ahuja}, \bibinfo{person}{Samarth Sikand}, \bibinfo{person}{Raghotham~M. Rao},
  \bibinfo{person}{R.~P. Jagadeesh~Chandra Bose}, \bibinfo{person}{Neville
  Dubash}, {and} \bibinfo{person}{Sanjay Podder}.}
  \bibinfo{year}{2018}\natexlab{}.
\newblock \showarticletitle{Identifying implementation bugs in machine learning
  based image classifiers using metamorphic testing}. In
  \bibinfo{booktitle}{\emph{Proc. of the 27th {ACM} {SIGSOFT} International
  Symposium on Software Testing and Analysis, {ISSTA}}}.
  \bibinfo{publisher}{{ACM}}, \bibinfo{pages}{118--128}.
\newblock
\urldef\tempurl%
\url{https://doi.org/10.1145/3213846.3213858}
\showDOI{\tempurl}


\bibitem[\protect\citeauthoryear{Elsken, Metzen, and Hutter}{Elsken
  et~al\mbox{.}}{2019}]%
        {DBLP:journals/jmlr/ElskenMH19}
\bibfield{author}{\bibinfo{person}{Thomas Elsken}, \bibinfo{person}{Jan~Hendrik
  Metzen}, {and} \bibinfo{person}{Frank Hutter}.}
  \bibinfo{year}{2019}\natexlab{}.
\newblock \showarticletitle{Neural Architecture Search: {A} Survey}.
\newblock \bibinfo{journal}{\emph{J. Mach. Learn. Res.}}  \bibinfo{volume}{20}
  (\bibinfo{year}{2019}), \bibinfo{pages}{55:1--55:21}.
\newblock


\bibitem[\protect\citeauthoryear{Fogel}{Fogel}{1997}]%
        {DBLP:journals/complexity/Fogel97}
\bibfield{author}{\bibinfo{person}{David~B. Fogel}.}
  \bibinfo{year}{1997}\natexlab{}.
\newblock \showarticletitle{Evolutionary algorithms in theory and practice}.
\newblock \bibinfo{journal}{\emph{Complex.}} \bibinfo{volume}{2},
  \bibinfo{number}{4} (\bibinfo{year}{1997}), \bibinfo{pages}{26--27}.
\newblock
\urldef\tempurl%
\url{https://doi.org/10.1002/(SICI)1099-0526(199703/04)2:4<26::AID-CPLX6>3.0.CO;2-7}
\showDOI{\tempurl}


\bibitem[\protect\citeauthoryear{Goldberg}{Goldberg}{1991}]%
        {DBLP:journals/csur/Goldberg91}
\bibfield{author}{\bibinfo{person}{David Goldberg}.}
  \bibinfo{year}{1991}\natexlab{}.
\newblock \showarticletitle{What Every Computer Scientist Should Know About
  Floating-Point Arithmetic}.
\newblock \bibinfo{journal}{\emph{{ACM} Comput. Surv.}} \bibinfo{volume}{23},
  \bibinfo{number}{1} (\bibinfo{year}{1991}), \bibinfo{pages}{5--48}.
\newblock
\urldef\tempurl%
\url{https://doi.org/10.1145/103162.103163}
\showDOI{\tempurl}


\bibitem[\protect\citeauthoryear{Goodfellow, Bengio, and Courville}{Goodfellow
  et~al\mbox{.}}{2016}]%
        {DBLP:books/daglib/0040158}
\bibfield{author}{\bibinfo{person}{Ian~J. Goodfellow}, \bibinfo{person}{Yoshua
  Bengio}, {and} \bibinfo{person}{Aaron~C. Courville}.}
  \bibinfo{year}{2016}\natexlab{}.
\newblock \bibinfo{booktitle}{\emph{Deep Learning}}.
\newblock \bibinfo{publisher}{{MIT} Press}.
\newblock
\showISBNx{978-0-262-03561-3}
\urldef\tempurl%
\url{http://www.deeplearningbook.org/}
\showURL{%
\tempurl}


\bibitem[\protect\citeauthoryear{Goodfellow, Pouget{-}Abadie, Mirza, Xu,
  Warde{-}Farley, Ozair, Courville, and Bengio}{Goodfellow
  et~al\mbox{.}}{2014}]%
        {DBLP:conf/nips/GoodfellowPMXWOCB14}
\bibfield{author}{\bibinfo{person}{Ian~J. Goodfellow}, \bibinfo{person}{Jean
  Pouget{-}Abadie}, \bibinfo{person}{Mehdi Mirza}, \bibinfo{person}{Bing Xu},
  \bibinfo{person}{David Warde{-}Farley}, \bibinfo{person}{Sherjil Ozair},
  \bibinfo{person}{Aaron~C. Courville}, {and} \bibinfo{person}{Yoshua Bengio}.}
  \bibinfo{year}{2014}\natexlab{}.
\newblock \showarticletitle{Generative Adversarial Nets}. In
  \bibinfo{booktitle}{\emph{Advances in Neural Information Processing Systems
  27: Annual Conference on Neural Information Processing Systems}}.
  \bibinfo{pages}{2672--2680}.
\newblock


\bibitem[\protect\citeauthoryear{Gu, Wen, Wang, Zhao, Luo, Kang, Zhou, Yang,
  Sun, Xu, Qiao, Li, Lin, and Zhang}{Gu et~al\mbox{.}}{2020}]%
        {DBLP:conf/sigsoft/GuWWZLKZYSXQLLZ20}
\bibfield{author}{\bibinfo{person}{Jiazhen Gu}, \bibinfo{person}{Jiaqi Wen},
  \bibinfo{person}{Zijian Wang}, \bibinfo{person}{Pu Zhao},
  \bibinfo{person}{Chuan Luo}, \bibinfo{person}{Yu Kang},
  \bibinfo{person}{Yangfan Zhou}, \bibinfo{person}{Li Yang},
  \bibinfo{person}{Jeffrey Sun}, \bibinfo{person}{Zhangwei Xu},
  \bibinfo{person}{Bo Qiao}, \bibinfo{person}{Liqun Li},
  \bibinfo{person}{Qingwei Lin}, {and} \bibinfo{person}{Dongmei Zhang}.}
  \bibinfo{year}{2020}\natexlab{}.
\newblock \showarticletitle{Efficient customer incident triage via linking with
  system incidents}. In \bibinfo{booktitle}{\emph{Proc. of the 28th {ACM} Joint
  European Software Engineering Conference and Symposium on the Foundations of
  Software Engineering, {ESEC/FSE}}}. \bibinfo{publisher}{{ACM}},
  \bibinfo{pages}{1296--1307}.
\newblock
\urldef\tempurl%
\url{https://doi.org/10.1145/3368089.3417061}
\showDOI{\tempurl}


\bibitem[\protect\citeauthoryear{Gulzar, Zhu, and Han}{Gulzar
  et~al\mbox{.}}{2019}]%
        {DBLP:conf/icse/GulzarZH19}
\bibfield{author}{\bibinfo{person}{Muhammad~Ali Gulzar},
  \bibinfo{person}{Yongkang Zhu}, {and} \bibinfo{person}{Xiaofeng Han}.}
  \bibinfo{year}{2019}\natexlab{}.
\newblock \showarticletitle{Perception and practices of differential testing}.
  In \bibinfo{booktitle}{\emph{Proc. of the 41st International Conference on
  Software Engineering: Software Engineering in Practice, {ICSE} {(SEIP)}}}.
  \bibinfo{publisher}{{IEEE} / {ACM}}, \bibinfo{pages}{71--80}.
\newblock
\urldef\tempurl%
\url{https://doi.org/10.1109/ICSE-SEIP.2019.00016}
\showDOI{\tempurl}


\bibitem[\protect\citeauthoryear{Guo, Jiang, Zhao, Chen, and Sun}{Guo
  et~al\mbox{.}}{2018}]%
        {DBLP:conf/sigsoft/GuoJZCS18}
\bibfield{author}{\bibinfo{person}{Jianmin Guo}, \bibinfo{person}{Yu Jiang},
  \bibinfo{person}{Yue Zhao}, \bibinfo{person}{Quan Chen}, {and}
  \bibinfo{person}{Jiaguang Sun}.} \bibinfo{year}{2018}\natexlab{}.
\newblock \showarticletitle{DLFuzz: differential fuzzing testing of deep
  learning systems}. In \bibinfo{booktitle}{\emph{Proc. of the 2018 {ACM} Joint
  Meeting on European Software Engineering Conference and Symposium on the
  Foundations of Software Engineering, {ESEC/FSE}}}.
  \bibinfo{publisher}{{ACM}}, \bibinfo{pages}{739--743}.
\newblock
\urldef\tempurl%
\url{https://doi.org/10.1145/3236024.3264835}
\showDOI{\tempurl}


\bibitem[\protect\citeauthoryear{Gupta, Anpalagan, Guan, and Khwaja}{Gupta
  et~al\mbox{.}}{2021}]%
        {DBLP:journals/array/GuptaAGK21}
\bibfield{author}{\bibinfo{person}{Abhishek Gupta}, \bibinfo{person}{Alagan
  Anpalagan}, \bibinfo{person}{Ling Guan}, {and}
  \bibinfo{person}{Ahmed~Shaharyar Khwaja}.} \bibinfo{year}{2021}\natexlab{}.
\newblock \showarticletitle{Deep learning for object detection and scene
  perception in self-driving cars: Survey, challenges, and open issues}.
\newblock \bibinfo{journal}{\emph{Array}}  \bibinfo{volume}{10}
  (\bibinfo{year}{2021}), \bibinfo{pages}{100057}.
\newblock
\urldef\tempurl%
\url{https://doi.org/10.1016/j.array.2021.100057}
\showDOI{\tempurl}


\bibitem[\protect\citeauthoryear{He, Zhang, Ren, and Sun}{He
  et~al\mbox{.}}{2016}]%
        {DBLP:conf/cvpr/HeZRS16}
\bibfield{author}{\bibinfo{person}{Kaiming He}, \bibinfo{person}{Xiangyu
  Zhang}, \bibinfo{person}{Shaoqing Ren}, {and} \bibinfo{person}{Jian Sun}.}
  \bibinfo{year}{2016}\natexlab{}.
\newblock \showarticletitle{Deep Residual Learning for Image Recognition}. In
  \bibinfo{booktitle}{\emph{2016 {IEEE} Conference on Computer Vision and
  Pattern Recognition, {CVPR}}}. \bibinfo{publisher}{{IEEE} Computer Society},
  \bibinfo{pages}{770--778}.
\newblock
\urldef\tempurl%
\url{https://doi.org/10.1109/CVPR.2016.90}
\showDOI{\tempurl}


\bibitem[\protect\citeauthoryear{Hecht{-}Nielsen}{Hecht{-}Nielsen}{1988}]%
        {DBLP:journals/nn/Hecht-Nielsen88a}
\bibfield{author}{\bibinfo{person}{Robert Hecht{-}Nielsen}.}
  \bibinfo{year}{1988}\natexlab{}.
\newblock \showarticletitle{Theory of the backpropagation neural network}.
\newblock \bibinfo{journal}{\emph{Neural Networks}} \bibinfo{volume}{1},
  \bibinfo{number}{Supplement-1} (\bibinfo{year}{1988}),
  \bibinfo{pages}{445--448}.
\newblock
\urldef\tempurl%
\url{https://doi.org/10.1016/0893-6080(88)90469-8}
\showDOI{\tempurl}


\bibitem[\protect\citeauthoryear{Huang, Liu, van~der Maaten, and
  Weinberger}{Huang et~al\mbox{.}}{2017}]%
        {DBLP:conf/cvpr/HuangLMW17}
\bibfield{author}{\bibinfo{person}{Gao Huang}, \bibinfo{person}{Zhuang Liu},
  \bibinfo{person}{Laurens van~der Maaten}, {and} \bibinfo{person}{Kilian~Q.
  Weinberger}.} \bibinfo{year}{2017}\natexlab{}.
\newblock \showarticletitle{Densely Connected Convolutional Networks}. In
  \bibinfo{booktitle}{\emph{2017 {IEEE} Conference on Computer Vision and
  Pattern Recognition, {CVPR}}}. \bibinfo{publisher}{{IEEE} Computer Society},
  \bibinfo{pages}{2261--2269}.
\newblock
\urldef\tempurl%
\url{https://doi.org/10.1109/CVPR.2017.243}
\showDOI{\tempurl}


\bibitem[\protect\citeauthoryear{Hung, Goodman, Ravel, Lopes, Rangel, Nery,
  Malleret, Nosten, Lacerda, Ferreira, R{\'{e}}nia, Duraisingh, Costa, Marti,
  and Carpenter}{Hung et~al\mbox{.}}{2020}]%
        {DBLP:journals/bmcbi/HungGRLRNMNLFRD20}
\bibfield{author}{\bibinfo{person}{Jane Hung}, \bibinfo{person}{Allen Goodman},
  \bibinfo{person}{Deepali Ravel}, \bibinfo{person}{Stefanie Lopes},
  \bibinfo{person}{Gabriel Rangel}, \bibinfo{person}{Odailton~A. Nery},
  \bibinfo{person}{Benoit Malleret}, \bibinfo{person}{Francois Nosten},
  \bibinfo{person}{Marcus V.~G. Lacerda}, \bibinfo{person}{Marcelo~U.
  Ferreira}, \bibinfo{person}{Laurent R{\'{e}}nia}, \bibinfo{person}{Manoj
  Duraisingh}, \bibinfo{person}{Fabio T.~M. Costa}, \bibinfo{person}{Matthias
  Marti}, {and} \bibinfo{person}{Anne~E. Carpenter}.}
  \bibinfo{year}{2020}\natexlab{}.
\newblock \showarticletitle{Keras {R-CNN:} library for cell detection in
  biological images using deep neural networks}.
\newblock \bibinfo{journal}{\emph{{BMC} Bioinform.}} \bibinfo{volume}{21},
  \bibinfo{number}{1} (\bibinfo{year}{2020}), \bibinfo{pages}{300}.
\newblock
\urldef\tempurl%
\url{https://doi.org/10.1186/s12859-020-03635-x}
\showDOI{\tempurl}


\bibitem[\protect\citeauthoryear{Islam, Nguyen, Pan, and Rajan}{Islam
  et~al\mbox{.}}{2019}]%
        {DBLP:conf/sigsoft/IslamNPR19}
\bibfield{author}{\bibinfo{person}{Md~Johirul Islam}, \bibinfo{person}{Giang
  Nguyen}, \bibinfo{person}{Rangeet Pan}, {and} \bibinfo{person}{Hridesh
  Rajan}.} \bibinfo{year}{2019}\natexlab{}.
\newblock \showarticletitle{A comprehensive study on deep learning bug
  characteristics}. In \bibinfo{booktitle}{\emph{Proc. of the {ACM} Joint
  Meeting on European Software Engineering Conference and Symposium on the
  Foundations of Software Engineering, {ESEC/FSE}}}.
  \bibinfo{publisher}{{ACM}}, \bibinfo{pages}{510--520}.
\newblock
\urldef\tempurl%
\url{https://doi.org/10.1145/3338906.3338955}
\showDOI{\tempurl}


\bibitem[\protect\citeauthoryear{Kepuska and Bohouta}{Kepuska and
  Bohouta}{2018}]%
        {DBLP:conf/ccwc/KepuskaB18}
\bibfield{author}{\bibinfo{person}{Veton Kepuska} {and} \bibinfo{person}{Gamal
  Bohouta}.} \bibinfo{year}{2018}\natexlab{}.
\newblock \showarticletitle{Next-generation of virtual personal assistants
  (Microsoft Cortana, Apple Siri, Amazon Alexa and Google Home)}. In
  \bibinfo{booktitle}{\emph{{IEEE} 8th Annual Computing and Communication
  Workshop and Conference, {CCWC}}}. \bibinfo{publisher}{{IEEE}},
  \bibinfo{pages}{99--103}.
\newblock
\urldef\tempurl%
\url{https://doi.org/10.1109/CCWC.2018.8301638}
\showDOI{\tempurl}


\bibitem[\protect\citeauthoryear{Kim, Feldt, and Yoo}{Kim
  et~al\mbox{.}}{2019}]%
        {DBLP:conf/icse/KimFY19}
\bibfield{author}{\bibinfo{person}{Jinhan Kim}, \bibinfo{person}{Robert Feldt},
  {and} \bibinfo{person}{Shin Yoo}.} \bibinfo{year}{2019}\natexlab{}.
\newblock \showarticletitle{Guiding deep learning system testing using surprise
  adequacy}. In \bibinfo{booktitle}{\emph{Proc. of the 41st International
  Conference on Software Engineering,{ICSE}}}. \bibinfo{publisher}{{IEEE} /
  {ACM}}, \bibinfo{pages}{1039--1049}.
\newblock
\urldef\tempurl%
\url{https://doi.org/10.1109/ICSE.2019.00108}
\showDOI{\tempurl}


\bibitem[\protect\citeauthoryear{Ma, Juefei{-}Xu, Zhang, Sun, Xue, Li, Chen,
  Su, Li, Liu, Zhao, and Wang}{Ma et~al\mbox{.}}{2018a}]%
        {DBLP:conf/kbse/MaJZSXLCSLLZW18}
\bibfield{author}{\bibinfo{person}{Lei Ma}, \bibinfo{person}{Felix
  Juefei{-}Xu}, \bibinfo{person}{Fuyuan Zhang}, \bibinfo{person}{Jiyuan Sun},
  \bibinfo{person}{Minhui Xue}, \bibinfo{person}{Bo Li},
  \bibinfo{person}{Chunyang Chen}, \bibinfo{person}{Ting Su},
  \bibinfo{person}{Li Li}, \bibinfo{person}{Yang Liu}, \bibinfo{person}{Jianjun
  Zhao}, {and} \bibinfo{person}{Yadong Wang}.}
  \bibinfo{year}{2018}\natexlab{a}.
\newblock \showarticletitle{DeepGauge: multi-granularity testing criteria for
  deep learning systems}. In \bibinfo{booktitle}{\emph{Proc. of the 33rd
  {ACM/IEEE} International Conference on Automated Software Engineering,
  {ASE}}}. \bibinfo{publisher}{{ACM}}, \bibinfo{pages}{120--131}.
\newblock
\urldef\tempurl%
\url{https://doi.org/10.1145/3238147.3238202}
\showDOI{\tempurl}


\bibitem[\protect\citeauthoryear{Ma, Zhang, Sun, Xue, Li, Juefei{-}Xu, Xie, Li,
  Liu, Zhao, and Wang}{Ma et~al\mbox{.}}{2018c}]%
        {DBLP:conf/issre/MaZSXLJXLLZW18}
\bibfield{author}{\bibinfo{person}{Lei Ma}, \bibinfo{person}{Fuyuan Zhang},
  \bibinfo{person}{Jiyuan Sun}, \bibinfo{person}{Minhui Xue},
  \bibinfo{person}{Bo Li}, \bibinfo{person}{Felix Juefei{-}Xu},
  \bibinfo{person}{Chao Xie}, \bibinfo{person}{Li Li}, \bibinfo{person}{Yang
  Liu}, \bibinfo{person}{Jianjun Zhao}, {and} \bibinfo{person}{Yadong Wang}.}
  \bibinfo{year}{2018}\natexlab{c}.
\newblock \showarticletitle{DeepMutation: Mutation Testing of Deep Learning
  Systems}. In \bibinfo{booktitle}{\emph{Proc. of the 29th {IEEE} International
  Symposium on Software Reliability Engineering, {ISSRE}}}.
  \bibinfo{publisher}{{IEEE} Computer Society}, \bibinfo{pages}{100--111}.
\newblock
\urldef\tempurl%
\url{https://doi.org/10.1109/ISSRE.2018.00021}
\showDOI{\tempurl}


\bibitem[\protect\citeauthoryear{Ma, Liu, Lee, Zhang, and Grama}{Ma
  et~al\mbox{.}}{2018b}]%
        {DBLP:conf/sigsoft/MaLLZG18}
\bibfield{author}{\bibinfo{person}{Shiqing Ma}, \bibinfo{person}{Yingqi Liu},
  \bibinfo{person}{Wen{-}Chuan Lee}, \bibinfo{person}{Xiangyu Zhang}, {and}
  \bibinfo{person}{Ananth Grama}.} \bibinfo{year}{2018}\natexlab{b}.
\newblock \showarticletitle{{MODE:} automated neural network model debugging
  via state differential analysis and input selection}. In
  \bibinfo{booktitle}{\emph{Proc. of the 2018 {ACM} Joint Meeting on European
  Software Engineering Conference and Symposium on the Foundations of Software
  Engineering, {ESEC/FSE}}}. \bibinfo{publisher}{{ACM}},
  \bibinfo{pages}{175--186}.
\newblock
\urldef\tempurl%
\url{https://doi.org/10.1145/3236024.3236082}
\showDOI{\tempurl}


\bibitem[\protect\citeauthoryear{Mikolov, Karafi{\'{a}}t, Burget,
  Cernock{\'{y}}, and Khudanpur}{Mikolov et~al\mbox{.}}{2010}]%
        {DBLP:conf/interspeech/MikolovKBCK10}
\bibfield{author}{\bibinfo{person}{Tom{\'{a}}s Mikolov},
  \bibinfo{person}{Martin Karafi{\'{a}}t}, \bibinfo{person}{Luk{\'{a}}s
  Burget}, \bibinfo{person}{Jan Cernock{\'{y}}}, {and} \bibinfo{person}{Sanjeev
  Khudanpur}.} \bibinfo{year}{2010}\natexlab{}.
\newblock \showarticletitle{Recurrent neural network based language model}. In
  \bibinfo{booktitle}{\emph{Proc. of the 11th Annual Conference of the
  International Speech Communication Association, {INTERSPEECH}}}.
  \bibinfo{publisher}{{ISCA}}, \bibinfo{pages}{1045--1048}.
\newblock


\bibitem[\protect\citeauthoryear{Muralikrishna, Vieira, dos Santos, and
  Almeida}{Muralikrishna et~al\mbox{.}}{2020}]%
        {DBLP:conf/iccsa/MuralikrishnaVS20}
\bibfield{author}{\bibinfo{person}{Amita Muralikrishna},
  \bibinfo{person}{Lu{\'{\i}}s Eduardo~Antunes Vieira}, \bibinfo{person}{Rafael
  Duarte~Coelho dos Santos}, {and} \bibinfo{person}{Adriano~P. Almeida}.}
  \bibinfo{year}{2020}\natexlab{}.
\newblock \showarticletitle{Total Solar Irradiance Forecasting with Keras
  Recurrent Neural Networks}. In \bibinfo{booktitle}{\emph{20th International
  Conference on Computational Science and Its Applications - {ICCSA}}}
  \emph{(\bibinfo{series}{Lecture Notes in Computer Science},
  Vol.~\bibinfo{volume}{12253})}. \bibinfo{publisher}{Springer},
  \bibinfo{pages}{255--269}.
\newblock
\urldef\tempurl%
\url{https://doi.org/10.1007/978-3-030-58814-4_18}
\showDOI{\tempurl}


\bibitem[\protect\citeauthoryear{Nair and Hinton}{Nair and Hinton}{2010}]%
        {DBLP:conf/icml/NairH10}
\bibfield{author}{\bibinfo{person}{Vinod Nair} {and}
  \bibinfo{person}{Geoffrey~E. Hinton}.} \bibinfo{year}{2010}\natexlab{}.
\newblock \showarticletitle{Rectified Linear Units Improve Restricted Boltzmann
  Machines}. In \bibinfo{booktitle}{\emph{Proc. of the 27th International
  Conference on Machine Learning, {ICML}}}. \bibinfo{publisher}{Omnipress},
  \bibinfo{pages}{807--814}.
\newblock


\bibitem[\protect\citeauthoryear{Odena, Olsson, Andersen, and Goodfellow}{Odena
  et~al\mbox{.}}{2019}]%
        {DBLP:conf/icml/OdenaOAG19}
\bibfield{author}{\bibinfo{person}{Augustus Odena}, \bibinfo{person}{Catherine
  Olsson}, \bibinfo{person}{David~G. Andersen}, {and} \bibinfo{person}{Ian~J.
  Goodfellow}.} \bibinfo{year}{2019}\natexlab{}.
\newblock \showarticletitle{TensorFuzz: Debugging Neural Networks with
  Coverage-Guided Fuzzing}. In \bibinfo{booktitle}{\emph{Proc. of the 36th
  International Conference on Machine Learning, {ICML}}}
  \emph{(\bibinfo{series}{Proceedings of Machine Learning Research},
  Vol.~\bibinfo{volume}{97})}. \bibinfo{publisher}{{PMLR}},
  \bibinfo{pages}{4901--4911}.
\newblock


\bibitem[\protect\citeauthoryear{Pei, Cao, Yang, and Jana}{Pei
  et~al\mbox{.}}{2017}]%
        {DBLP:conf/sosp/PeiCYJ17}
\bibfield{author}{\bibinfo{person}{Kexin Pei}, \bibinfo{person}{Yinzhi Cao},
  \bibinfo{person}{Junfeng Yang}, {and} \bibinfo{person}{Suman Jana}.}
  \bibinfo{year}{2017}\natexlab{}.
\newblock \showarticletitle{DeepXplore: Automated Whitebox Testing of Deep
  Learning Systems}. In \bibinfo{booktitle}{\emph{Proc. of the 26th Symposium
  on Operating Systems Principles, {SOSP}}}. \bibinfo{publisher}{{ACM}},
  \bibinfo{pages}{1--18}.
\newblock
\urldef\tempurl%
\url{https://doi.org/10.1145/3132747.3132785}
\showDOI{\tempurl}


\bibitem[\protect\citeauthoryear{Pham, Lutellier, Qi, and Tan}{Pham
  et~al\mbox{.}}{2019}]%
        {DBLP:conf/icse/PhamLQT19}
\bibfield{author}{\bibinfo{person}{Hung~Viet Pham}, \bibinfo{person}{Thibaud
  Lutellier}, \bibinfo{person}{Weizhen Qi}, {and} \bibinfo{person}{Lin Tan}.}
  \bibinfo{year}{2019}\natexlab{}.
\newblock \showarticletitle{{CRADLE:} cross-backend validation to detect and
  localize bugs in deep learning libraries}. In \bibinfo{booktitle}{\emph{Proc.
  of the 41st International Conference on Software Engineering,{ICSE}}}.
  \bibinfo{publisher}{{IEEE} / {ACM}}, \bibinfo{pages}{1027--1038}.
\newblock
\urldef\tempurl%
\url{https://doi.org/10.1109/ICSE.2019.00107}
\showDOI{\tempurl}


\bibitem[\protect\citeauthoryear{Simonyan and Zisserman}{Simonyan and
  Zisserman}{2015}]%
        {DBLP:journals/corr/SimonyanZ14a}
\bibfield{author}{\bibinfo{person}{Karen Simonyan} {and}
  \bibinfo{person}{Andrew Zisserman}.} \bibinfo{year}{2015}\natexlab{}.
\newblock \showarticletitle{Very Deep Convolutional Networks for Large-Scale
  Image Recognition}. In \bibinfo{booktitle}{\emph{Proc. of the 3rd
  International Conference on Learning Representations, {ICLR}}},
  \bibfield{editor}{\bibinfo{person}{Yoshua Bengio} {and} \bibinfo{person}{Yann
  LeCun}} (Eds.).
\newblock


\bibitem[\protect\citeauthoryear{Srivastava, Bond, McKinley, and
  Shmatikov}{Srivastava et~al\mbox{.}}{2011}]%
        {DBLP:conf/pldi/SrivastavaBMS11}
\bibfield{author}{\bibinfo{person}{Varun Srivastava},
  \bibinfo{person}{Michael~D. Bond}, \bibinfo{person}{Kathryn~S. McKinley},
  {and} \bibinfo{person}{Vitaly Shmatikov}.} \bibinfo{year}{2011}\natexlab{}.
\newblock \showarticletitle{A security policy oracle: detecting security holes
  using multiple {API} implementations}. In \bibinfo{booktitle}{\emph{Proc. of
  the 32nd {ACM} {SIGPLAN} Conference on Programming Language Design and
  Implementation, {PLDI}}}. \bibinfo{publisher}{{ACM}},
  \bibinfo{pages}{343--354}.
\newblock
\urldef\tempurl%
\url{https://doi.org/10.1145/1993316.1993539}
\showDOI{\tempurl}


\bibitem[\protect\citeauthoryear{Team, Al-Rfou, Alain, Almahairi, Angermueller,
  Bahdanau, Ballas, Bastien, Bayer, Belikov, et~al\mbox{.}}{Team
  et~al\mbox{.}}{2016}]%
        {team2016theano}
\bibfield{author}{\bibinfo{person}{The Theano~Development Team},
  \bibinfo{person}{Rami Al-Rfou}, \bibinfo{person}{Guillaume Alain},
  \bibinfo{person}{Amjad Almahairi}, \bibinfo{person}{Christof Angermueller},
  \bibinfo{person}{Dzmitry Bahdanau}, \bibinfo{person}{Nicolas Ballas},
  \bibinfo{person}{Fr{\'e}d{\'e}ric Bastien}, \bibinfo{person}{Justin Bayer},
  \bibinfo{person}{Anatoly Belikov}, {et~al\mbox{.}}}
  \bibinfo{year}{2016}\natexlab{}.
\newblock \showarticletitle{Theano: A Python framework for fast computation of
  mathematical expressions}.
\newblock \bibinfo{journal}{\emph{arXiv preprint arXiv:1605.02688}}
  (\bibinfo{year}{2016}).
\newblock


\bibitem[\protect\citeauthoryear{Tian, Pei, Jana, and Ray}{Tian
  et~al\mbox{.}}{2018}]%
        {DBLP:conf/icse/TianPJR18}
\bibfield{author}{\bibinfo{person}{Yuchi Tian}, \bibinfo{person}{Kexin Pei},
  \bibinfo{person}{Suman Jana}, {and} \bibinfo{person}{Baishakhi Ray}.}
  \bibinfo{year}{2018}\natexlab{}.
\newblock \showarticletitle{DeepTest: automated testing of
  deep-neural-network-driven autonomous cars}. In
  \bibinfo{booktitle}{\emph{Proc. of the 40th International Conference on
  Software Engineering,{ICSE}}}. \bibinfo{publisher}{{ACM}},
  \bibinfo{pages}{303--314}.
\newblock
\urldef\tempurl%
\url{https://doi.org/10.1145/3180155.3180220}
\showDOI{\tempurl}


\bibitem[\protect\citeauthoryear{Wang, Li, Xiao, Zhu, Li, Wong, and Chao}{Wang
  et~al\mbox{.}}{2019}]%
        {DBLP:conf/acl/WangLXZLWC19}
\bibfield{author}{\bibinfo{person}{Qiang Wang}, \bibinfo{person}{Bei Li},
  \bibinfo{person}{Tong Xiao}, \bibinfo{person}{Jingbo Zhu},
  \bibinfo{person}{Changliang Li}, \bibinfo{person}{Derek~F. Wong}, {and}
  \bibinfo{person}{Lidia~S. Chao}.} \bibinfo{year}{2019}\natexlab{}.
\newblock \showarticletitle{Learning Deep Transformer Models for Machine
  Translation}. In \bibinfo{booktitle}{\emph{Proc. of the 57th Conference of
  the Association for Computational Linguistics, {ACL}, Volume 1: Long
  Papers}}. \bibinfo{publisher}{Association for Computational Linguistics},
  \bibinfo{pages}{1810--1822}.
\newblock
\urldef\tempurl%
\url{https://doi.org/10.18653/v1/P19-1176}
\showDOI{\tempurl}


\bibitem[\protect\citeauthoryear{Wang, Yan, Chen, Liu, and Zhang}{Wang
  et~al\mbox{.}}{2020}]%
        {DBLP:conf/sigsoft/WangYCLZ20}
\bibfield{author}{\bibinfo{person}{Zan Wang}, \bibinfo{person}{Ming Yan},
  \bibinfo{person}{Junjie Chen}, \bibinfo{person}{Shuang Liu}, {and}
  \bibinfo{person}{Dongdi Zhang}.} \bibinfo{year}{2020}\natexlab{}.
\newblock \showarticletitle{Deep learning library testing via effective model
  generation}. In \bibinfo{booktitle}{\emph{Proc. of the 28th {ACM} Joint
  European Software Engineering Conference and Symposium on the Foundations of
  Software Engineering, {ESEC/FSE}}}. \bibinfo{publisher}{{ACM}},
  \bibinfo{pages}{788--799}.
\newblock
\urldef\tempurl%
\url{https://doi.org/10.1145/3368089.3409761}
\showDOI{\tempurl}


\bibitem[\protect\citeauthoryear{Wardat, Le, and Rajan}{Wardat
  et~al\mbox{.}}{2021}]%
        {DBLP:conf/icse/WardatLR21}
\bibfield{author}{\bibinfo{person}{Mohammad Wardat}, \bibinfo{person}{Wei Le},
  {and} \bibinfo{person}{Hridesh Rajan}.} \bibinfo{year}{2021}\natexlab{}.
\newblock \showarticletitle{DeepLocalize: Fault Localization for Deep Neural
  Networks}. In \bibinfo{booktitle}{\emph{Proc.of the 43rd {IEEE/ACM}
  International Conference on Software Engineering, {ICSE}}}.
  \bibinfo{publisher}{{IEEE}}, \bibinfo{pages}{251--262}.
\newblock
\urldef\tempurl%
\url{https://doi.org/10.1109/ICSE43902.2021.00034}
\showDOI{\tempurl}


\bibitem[\protect\citeauthoryear{Wistuba, Rawat, and Pedapati}{Wistuba
  et~al\mbox{.}}{2019}]%
        {DBLP:journals/corr/abs-1905-01392}
\bibfield{author}{\bibinfo{person}{Martin Wistuba}, \bibinfo{person}{Ambrish
  Rawat}, {and} \bibinfo{person}{Tejaswini Pedapati}.}
  \bibinfo{year}{2019}\natexlab{}.
\newblock \showarticletitle{A Survey on Neural Architecture Search}.
\newblock \bibinfo{journal}{\emph{CoRR}}  \bibinfo{volume}{abs/1905.01392}
  (\bibinfo{year}{2019}).
\newblock
\showeprint[arxiv]{1905.01392}
\urldef\tempurl%
\url{http://arxiv.org/abs/1905.01392}
\showURL{%
\tempurl}


\bibitem[\protect\citeauthoryear{Wu, Feichtenhofer, Fan, He,
  Kr{\"{a}}henb{\"{u}}hl, and Girshick}{Wu et~al\mbox{.}}{2019a}]%
        {DBLP:conf/cvpr/WuF0HKG19}
\bibfield{author}{\bibinfo{person}{Chao{-}Yuan Wu}, \bibinfo{person}{Christoph
  Feichtenhofer}, \bibinfo{person}{Haoqi Fan}, \bibinfo{person}{Kaiming He},
  \bibinfo{person}{Philipp Kr{\"{a}}henb{\"{u}}hl}, {and}
  \bibinfo{person}{Ross~B. Girshick}.} \bibinfo{year}{2019}\natexlab{a}.
\newblock \showarticletitle{Long-Term Feature Banks for Detailed Video
  Understanding}. In \bibinfo{booktitle}{\emph{{IEEE} Conference on Computer
  Vision and Pattern Recognition, {CVPR}}}. \bibinfo{publisher}{Computer Vision
  Foundation / {IEEE}}, \bibinfo{pages}{284--293}.
\newblock
\urldef\tempurl%
\url{https://doi.org/10.1109/CVPR.2019.00037}
\showDOI{\tempurl}


\bibitem[\protect\citeauthoryear{Wu, Xu, Zhong, Lyu, and King}{Wu
  et~al\mbox{.}}{2019b}]%
        {DBLP:conf/dsn/WuXZLK19}
\bibfield{author}{\bibinfo{person}{Weibin Wu}, \bibinfo{person}{Hui Xu},
  \bibinfo{person}{Sanqiang Zhong}, \bibinfo{person}{Michael~R. Lyu}, {and}
  \bibinfo{person}{Irwin King}.} \bibinfo{year}{2019}\natexlab{b}.
\newblock \showarticletitle{Deep Validation: Toward Detecting Real-World Corner
  Cases for Deep Neural Networks}. In \bibinfo{booktitle}{\emph{Proc. of the
  49th Annual {IEEE/IFIP} International Conference on Dependable Systems and
  Networks, {DSN}}}. \bibinfo{publisher}{{IEEE}}, \bibinfo{pages}{125--137}.
\newblock
\urldef\tempurl%
\url{https://doi.org/10.1109/DSN.2019.00026}
\showDOI{\tempurl}


\bibitem[\protect\citeauthoryear{Xie, Ma, Juefei{-}Xu, Xue, Chen, Liu, Zhao,
  Li, Yin, and See}{Xie et~al\mbox{.}}{2019}]%
        {DBLP:conf/issta/XieMJXCLZLYS19}
\bibfield{author}{\bibinfo{person}{Xiaofei Xie}, \bibinfo{person}{Lei Ma},
  \bibinfo{person}{Felix Juefei{-}Xu}, \bibinfo{person}{Minhui Xue},
  \bibinfo{person}{Hongxu Chen}, \bibinfo{person}{Yang Liu},
  \bibinfo{person}{Jianjun Zhao}, \bibinfo{person}{Bo Li},
  \bibinfo{person}{Jianxiong Yin}, {and} \bibinfo{person}{Simon See}.}
  \bibinfo{year}{2019}\natexlab{}.
\newblock \showarticletitle{DeepHunter: a coverage-guided fuzz testing
  framework for deep neural networks}. In \bibinfo{booktitle}{\emph{Proc. of
  the 28th {ACM} {SIGSOFT} International Symposium on Software Testing and
  Analysis, {ISSTA}}}. \bibinfo{publisher}{{ACM}}, \bibinfo{pages}{146--157}.
\newblock
\urldef\tempurl%
\url{https://doi.org/10.1145/3293882.3330579}
\showDOI{\tempurl}


\bibitem[\protect\citeauthoryear{Xie, Zhang, Chen, Liu, Poon, and Xu}{Xie
  et~al\mbox{.}}{2020}]%
        {DBLP:journals/tr/XieZCLPX20}
\bibfield{author}{\bibinfo{person}{Xiaoyuan Xie}, \bibinfo{person}{Zhiyi
  Zhang}, \bibinfo{person}{Tsong~Yueh Chen}, \bibinfo{person}{Yang Liu},
  \bibinfo{person}{Pak{-}Lok Poon}, {and} \bibinfo{person}{Baowen Xu}.}
  \bibinfo{year}{2020}\natexlab{}.
\newblock \showarticletitle{{METTLE:} {A} METamorphic Testing Approach to
  Assessing and Validating Unsupervised Machine Learning Systems}.
\newblock \bibinfo{journal}{\emph{{IEEE} Trans. Reliab.}} \bibinfo{volume}{69},
  \bibinfo{number}{4} (\bibinfo{year}{2020}), \bibinfo{pages}{1293--1322}.
\newblock
\urldef\tempurl%
\url{https://doi.org/10.1109/TR.2020.2972266}
\showDOI{\tempurl}


\bibitem[\protect\citeauthoryear{Xu, Wang, Chen, and Li}{Xu
  et~al\mbox{.}}{2015}]%
        {DBLP:journals/corr/XuWCL15}
\bibfield{author}{\bibinfo{person}{Bing Xu}, \bibinfo{person}{Naiyan Wang},
  \bibinfo{person}{Tianqi Chen}, {and} \bibinfo{person}{Mu Li}.}
  \bibinfo{year}{2015}\natexlab{}.
\newblock \showarticletitle{Empirical Evaluation of Rectified Activations in
  Convolutional Network}.
\newblock \bibinfo{journal}{\emph{CoRR}}  \bibinfo{volume}{abs/1505.00853}
  (\bibinfo{year}{2015}).
\newblock
\showeprint[arxiv]{1505.00853}


\bibitem[\protect\citeauthoryear{Yang, Chen, Eide, and Regehr}{Yang
  et~al\mbox{.}}{2011}]%
        {DBLP:conf/pldi/YangCER11}
\bibfield{author}{\bibinfo{person}{Xuejun Yang}, \bibinfo{person}{Yang Chen},
  \bibinfo{person}{Eric Eide}, {and} \bibinfo{person}{John Regehr}.}
  \bibinfo{year}{2011}\natexlab{}.
\newblock \showarticletitle{Finding and understanding bugs in {C} compilers}.
  In \bibinfo{booktitle}{\emph{Proc. of the 32nd {ACM} {SIGPLAN} Conference on
  Programming Language Design and Implementation, {PLDI}}}.
  \bibinfo{publisher}{{ACM}}, \bibinfo{pages}{283--294}.
\newblock
\urldef\tempurl%
\url{https://doi.org/10.1145/1993316.1993532}
\showDOI{\tempurl}


\bibitem[\protect\citeauthoryear{Zhang, Harman, Ma, and Liu}{Zhang
  et~al\mbox{.}}{2020a}]%
        {zhang2020machine}
\bibfield{author}{\bibinfo{person}{Jie~M Zhang}, \bibinfo{person}{Mark Harman},
  \bibinfo{person}{Lei Ma}, {and} \bibinfo{person}{Yang Liu}.}
  \bibinfo{year}{2020}\natexlab{a}.
\newblock \showarticletitle{Machine learning testing: Survey, landscapes and
  horizons}.
\newblock \bibinfo{journal}{\emph{IEEE Transactions on Software Engineering}}
  (\bibinfo{year}{2020}).
\newblock


\bibitem[\protect\citeauthoryear{Zhang, Zhang, Zhang, Liu, and Khurshid}{Zhang
  et~al\mbox{.}}{2018}]%
        {DBLP:conf/kbse/ZhangZZ0K18}
\bibfield{author}{\bibinfo{person}{Mengshi Zhang}, \bibinfo{person}{Yuqun
  Zhang}, \bibinfo{person}{Lingming Zhang}, \bibinfo{person}{Cong Liu}, {and}
  \bibinfo{person}{Sarfraz Khurshid}.} \bibinfo{year}{2018}\natexlab{}.
\newblock \showarticletitle{DeepRoad: GAN-based metamorphic testing and input
  validation framework for autonomous driving systems}. In
  \bibinfo{booktitle}{\emph{Proc. of the 33rd {ACM/IEEE} International
  Conference on Automated Software Engineering, {ASE}}}.
  \bibinfo{publisher}{{ACM}}, \bibinfo{pages}{132--142}.
\newblock
\urldef\tempurl%
\url{https://doi.org/10.1145/3238147.3238187}
\showDOI{\tempurl}


\bibitem[\protect\citeauthoryear{Zhang, Zhai, Ma, and Shen}{Zhang
  et~al\mbox{.}}{2021}]%
        {DBLP:conf/icse/ZhangZMS21}
\bibfield{author}{\bibinfo{person}{Xiaoyu Zhang}, \bibinfo{person}{Juan Zhai},
  \bibinfo{person}{Shiqing Ma}, {and} \bibinfo{person}{Chao Shen}.}
  \bibinfo{year}{2021}\natexlab{}.
\newblock \showarticletitle{{AUTOTRAINER:} An Automatic {DNN} Training Problem
  Detection and Repair System}. In \bibinfo{booktitle}{\emph{Proc. of the 43rd
  {IEEE/ACM} International Conference on Software Engineering,{ICSE}}}.
  \bibinfo{publisher}{{IEEE}}, \bibinfo{pages}{359--371}.
\newblock
\urldef\tempurl%
\url{https://doi.org/10.1109/ICSE43902.2021.00043}
\showDOI{\tempurl}


\bibitem[\protect\citeauthoryear{Zhang, Ren, Chen, Xiong, Cheung, and
  Xie}{Zhang et~al\mbox{.}}{2020b}]%
        {DBLP:conf/sigsoft/ZhangRC0C020}
\bibfield{author}{\bibinfo{person}{Yuhao Zhang}, \bibinfo{person}{Luyao Ren},
  \bibinfo{person}{Liqian Chen}, \bibinfo{person}{Yingfei Xiong},
  \bibinfo{person}{Shing{-}Chi Cheung}, {and} \bibinfo{person}{Tao Xie}.}
  \bibinfo{year}{2020}\natexlab{b}.
\newblock \showarticletitle{Detecting numerical bugs in neural network
  architectures}. In \bibinfo{booktitle}{\emph{Proc. of the 28th {ACM} Joint
  European Software Engineering Conference and Symposium on the Foundations of
  Software Engineering, {ESEC/FSE}}}. \bibinfo{publisher}{{ACM}},
  \bibinfo{pages}{826--837}.
\newblock
\urldef\tempurl%
\url{https://doi.org/10.1145/3368089.3409720}
\showDOI{\tempurl}


\end{thebibliography}
\balance


\end{document}